\documentclass[12pt]{iopart}

\usepackage{graphicx}
\usepackage{fancyhdr}
\usepackage{amsfonts,amssymb,color,amstext}
\usepackage{subfig}
\usepackage{float}
\usepackage{blindtext}

\definecolor{red}{rgb}{1,0,0}

\usepackage{pstool}
\def\nn{\nonumber}
\def\be{\begin{equation}}
\def\ee{\end{equation}}
\def\ba{\begin{eqnarray}}
\def\ea{\end{eqnarray}}

\begin{document}
\title[]{Quantisation ambiguities and the effective dynamics  of scalar-tensor theories in loop quantum cosmology}

\author{Yu Han}

\address{College of Physics and Electrical Engineering, Xinyang Normal University, Xinyang 464000, P.R. China}

\begin{abstract}	
The Hamiltonian constraint of scalar-tensor theories in the Jordan frame is quantised using three quantisation prescriptions in loop quantum cosmology,  from which we obtain  three different effective Hamiltonian constraints. The corresponding effective equations of motion derived from these effective Hamiltonian constraints turn out to be drastically different. The implications of each set of effective equations of motion are discussed in detail.
In the latter half of this paper, as a concrete example,  we study the effective dynamics of a specific model with the non-minimal coupling function $F(\phi)=1+\xi\kappa\phi^2$ and self-interacting quartic potential. Using numerical results, we find different features for different effective dynamics. Moreover, it is also found that the spacetime singularity is absent and the cosmological bounce exists in each effective dynamics of this model.
\end{abstract}

\maketitle


\section{Introduction}\label{sec1}

Constructing a consistent quantum theory of gravity is one of the most fundamental problems in theoretical physics. There have been a lot of tentative theories proposed to quantise gravity. One of the leading programs is loop quantum gravity (LQG), which has made much progress during the past few decades \cite{Ashtekar:2021,Ashtekar:2017}.

The essential feature of general relativity is interpreting the gravitational interaction as the geometric effect of spacetime. LQG takes this feature very seriously and believes that to quantise  gravity is just to quantise  the pseudo-Riemannian geometry. In the canonical formulation of LQG, the gravitational constraints are expressed in terms of the conjugate variables $\big(A_a^i,E^a_i\big)$, in which $A_a^i$ is the su(2)-valued Ashtekar connection and $E^a_i$ is the densitised triad which is an su(2)-valued vector density of weight one. Instead of directly quantising $\big(A_a^i,E^a_i\big)$, in LQG one quantises particular smeared versions of them, to be specific, the connection is integrated along an oriented curve $e$, yielding the associated holonomy $A(e)$, and the densitised triad is smeared over a two-dimensional surface $S$, yielding the associated flux $E(S)$, then, one can define the holonomy-flux algebra and finds suitable representations of it and derives the kinematical Hilbert space underlying the representation, after that, the gravitational constraints can be implemented as operators on the the kinematical Hilbert space, and the remaining tasks are to find physical solutions satisfying these quantum constraints \cite{Ashtekar:2004,Rovelli:2015}. However, it turns out to be extremely complicated and difficult to solve the quantum Hamiltonian constraint, and it seems that to attack this problem we must develop many new theoretical and especially numerical tools, which would be undoubtedly a long-term task. Under this situation, it is reasonable to first study some simple symmetry reduced systems using some techniques of LQG in order to obtain some preliminary information of quantum gravity from these simple systems.

Loop quantum cosmology (LQC) is a frame which aims to apply the quantisation methods of LQG to cosmological spacetimes \cite{Ashtekar:2015,Ashtekar:2011,Guillermo:2011,Bojowald:2005}. LQC is characterised by several quantum corrections different from other quantum cosmological theories, among which the most well studied is the holonomy correction \cite{Bojowald:2014}.
In the Hamiltonian constraint of the minimally coupled LQC, the curvature of the Ashtekar connection is replaced by functions of the holonomy of the connection.
 Several different types of holonomies have been used to regularise the curvature in the Hamiltonian constraint, one is to directly replace the curvature by the holonomy of the connection around closed loops (hereafter called the closed holonomy regularisation); another is to replace the connection  by its holonomy along open segments  and then use the regularised connection to express the curvature (hereafter called the open holonomy regularisation); the third is to use the holonomy of the extrinsic curvature (which can also be regarded as a connection variable in certain cases) along open segments to replace the extrinsic curvature itself and then use the regularised extrinsic curvature to express the curvature (hereafter called the K-holonomy regularisation). Since different spatial manifolds have different underlying symmetries, not all holonomy regularisation prescriptions are applicable to each model in LQC. To be specific, the applicability depends on whether the holonomy can be expressed in terms of almost periodic functions of the connection in this model, because to define the quantum Hamiltonian constraint in LQC one must first express the Hamiltonian constraint in terms of almost periodic functions of the connection which can be promoted to well defined operators. In the context of the minimally coupled LQC, the closed holonomy regularisation has been be applied to the $k=0$ and $k=1$ Friedmann-Robertson-Walker (FRW) models, the Bianchi-I model, the Gowdy models, etc. \cite{Ashtekar:2006a,Ashtekar:2006b,Ashtekar:2008,Szulc:2006,Ashtekar:2007,Ashtekar:2009a,Martin-Benito:2010}; the open holonomy regularisation has been be applied to the Bianchi-II and Bianchi-IX models and the $k=1$ FRW model  \cite{Ashtekar:2009b,Wilson-Ewing:2010,Corichi:2011}; the K-holonomy regularisation has been applied to the $k=-1$ and $k=1$ FRW models, the Bianchi-II and Bianchi-IX models \cite{Bojowald:2004,Vandersloot:2007,Singh:2014}. In the $k=0$ FRW model, the three holonomy regularisation prescriptions are equivalent.  In other cases, different choices of holonomy regularisations lead to different quantum and effective dynamics. In the quantum dynamics of all studied models in the minimally coupled LQC, the classical singularities are avoided and replaced by a quantum bounce caused by the holonomy corrections. It is worth mentioning that the existence of a quantum bounce has been verified in a quite different approach in LQG called the ``quantum reduced loop gravity" \cite{Alesci:2018}.  Hence, it is perhaps not inappropriate to say that although only finite degrees of freedom are quantized in LQC, the holonomy corrections do capture some essential features of the yet not very well understood dynamics of LQG.

A particular useful tool used in the minimally coupled LQC is the effective dynamics in which one can study the quantum evolution in a continuum spacetime manner at the the effective level through a modified differential Friedmann equation and Raychaudhuri equation which incorporate the leading quantum gravitational effects of LQC. These effective equations can be obtained from an effective Hamiltonian. The viability of the effective Hamiltonian is based on the fact that in LQC the states initially sharply peaked remain sharply peaked at all times, including the bounce point  and for sharply peaked states we can speak of a continuous spacetime on which dynamics is given by the effective Hamiltonian \cite{Corichi:2008,Kaminski:2010}. The effective Hamiltonian can be obtained following different routes in LQC \cite{Taveras:2007,Bojowald:2009,Ashtekar:2010}. Numerical simulations show that the full quantum dynamics and the effective dynamics generated by the effective Hamiltonian agree with each other very well \cite{Ashtekar:2006b,Diener:2014}.

Instead of the minimally coupled LQC, we are interested in the LQC of scalar-tensor theories (STT) in this paper.
Among modified gravity theories, STT have so far received more attention than others in the research of cosmology \cite{Shankaranarayanan:2022}.
In particular, the predictions of some very simple specific models of STT are in excellent agreement with the recent astrophysical observations for inflation \cite{Akrami:2018,Ade:2016}, which imply that STT might be the right alternative of general relativity in the very early universe. Therefore, it is necessary to study the quantum gravity effects of STT for future observations. It is well known that in the classical case the Einstein frame and the Jordan frame of STT can be related to each other by conformal transformation, or in other words, the Einstein frame and the Jordan frame of STT are mathematically equivalent. Using this equivalence, the classical STT are often studied in the simpler Einstein frame. In LQC, the holonomy corrections of some particular models of STT have also been studied in the Einstein frame \cite{Bojowald:2006a,Bojowald:2006b,Artymowski:2012,Amoros:2014,Odintsov:2014,Haro:2018,Li:2018a}, in which Einstein frame variables are regularised before quantisation; however, since the procedure of conformal transformation does not commute with the procedure of regularisation, the effective equations followed from quantising the Einstein frame variables are no longer mathematically equivalent to the effective equations followed from quantising the the Jordan frame variables. Hence, if we hold that the Jordan frame is the physical frame, the more reasonable way to study the quantum gravity effects of STT is to directly quantise the Jordan frame variables. Based on this belief, the connection dynamics of STT is formulated in the Jordan frame in  \cite{Zhang:2011a,Zhang:2011b}. The LQC of some specific models of STT are subsequently studied in the Jordan
  frame in
  \cite{Zhang:2013,Artymowski:2013,Jin:2018,Sharma:2019}, however, the effective equations derived in these references are incomplete, this crucial fact has been pointed out in \cite{Chen:2019,Han:2019} and the complete equations of motion are given there.  The implications of the effective equations of motion will be discussed in the current paper.

 The main focuses of the current paper are the quantisation ambiguities and the related effective dynamics of STT in the Jordan frame in the $k=0$ FRW model.  The structure of this paper is as follows. In section \ref{sec2},  we introduce the connection formulation of the classical Hamiltonian constraint of STT in the Jordan frame. In section \ref{sec3}, we quantise the Hamiltonian constraint using two different prescriptions, which leads to totally three different effective Hamiltonian constraints. Then, we derive the effective equations of motion from each effective Hamiltonian. Moreover, we also discuss the implications of each set of equations of motion and make a comparison between them.
In section \ref{sec4}, we use the results obtained in section \ref{sec3} to study the effective dynamics of a specific model of STT, which aims to illustrate the features of different quantisation prescriptions more concretely. In section \ref{sec5}, we draw the conclusions and make some remarks.

\section{Classical dynamics of STT}\label{sec2}

The action of STT we use in this paper reads
\ba
S=\int_{\Sigma}d^4x\sqrt{|\det(g)|}\Bigg[\frac{1}{2\kappa}F(\phi)R
-\frac{1}{2}K(\phi)(\partial^{\mu}\phi)\partial_{\mu}\phi-V(\phi)\Bigg],
\label{STTaction}
\ea
  in which $\Sigma$ is the spacetime manifold and $\kappa\equiv8\pi G$,  $F(\phi)$ and $K(\phi)$  are dimensionless coupling functions of the scalar field.

 In the canonical formulation of the theory, we introduce the su(2)-valued triad $e^a_i$ and its co-triad $\omega_a^i$ which are related to the ADM 3-metric by  $h_{ab}=\omega_a^i\omega_b^j\delta_{ij}$, $h^{ab}=e^a_ie^b_j\delta^{ij}$, using these variables we can define the densitised triad and the Ashtekar connection,
 \ba
 E^a_i:=\det(\omega)e^a_i,\qquad A_a^i:=\Gamma_a^i+\gamma K_a^i,
 \ea
  where $\Gamma_a^i$ is the spin connection compatible with the triad, $\gamma$ is the Barbero-Immirzi parameter and $K_a^i$ is related to the extrinsic curvature by $K_a^i\equiv \delta^{ij}K_{ab}e^b_j$.
   The elementary phase space variables are the geometric conjugate variables $\big(A_a^i,E^a_i\big)$ and the scalar conjugate variables $(\phi,\tilde{\pi})$, which satisfy
 \ba
 \big\{A_a^i(x),E^b_j(y)\big\}=\gamma\kappa\delta_j^i\delta^b_a\delta^{(3)}(x,y),
 \quad
 \big\{\phi(x),\tilde{\pi}(y)\big\}=\delta^{(3)}(x,y).
 \ea

 The Hamiltonian constraint of STT in the Jordan frame can be expressed as \cite{Han:2019,Han:2015}
 \ba
 \mathcal{C}&=&\frac{F(\phi)}{2\kappa\sqrt{|\det E}|}E^a_iE^b_j\Bigg[\epsilon^{ij}_{~~k}F^k_{ab}
 -2\Bigg(\gamma^2+\frac{1}{F^2(\phi)}\Bigg)K^i_{[a}K^j_{b]}\Bigg]\nn\\
 &&+\frac{1}{2F(\phi)G(\phi)\sqrt{|\det E|}}
 \Bigg[\frac{F'(\phi)}{\kappa}K_a^iE^a_i
 +F(\phi)\tilde{\pi}\Bigg]^2\nn\\
 &&+\sqrt{|\det E|}\Bigg[\frac{1}{\kappa}D^a D_a F(\phi )+\frac{K(\phi)}{2}(D^a\phi)D_a\phi+V(\phi)\Bigg]\nn\\
 &=&0,\label{newconstraint}
 \ea
 where $F^{i}_{ab}\equiv2\partial_{[a}A^i_{b]}+\epsilon^{~~i}_{jk}A^j_aA^k_b$ is the curvature of Ashtekar connection, and $G(\phi)$ is defined by
 \ba
 G(\phi):=\frac{3}{2\kappa}\big(F'(\phi)\big)^2+F(\phi)K(\phi),\label{Gphi}
 \ea
 in which the prime denotes the derivative with respective to $\phi$, i.e., $F'(\phi)\equiv\frac{dF(\phi)}{d\phi}$.
  In the $k=0$ FRW model, the line element of the spacetime metric is given by
  \ba
  ds^2=-N^2d\tau^2+a^2(\tau)\mathring{h}_{ab}dx^adx^b,\label{FRWmetric}
  \ea
where $N$ is the homogenous lapse function, $a$ is the scale factor, and $\mathring{h}_{ab}$ is some fiducial metric which is related to the physical metric by $h_{ab}=a^2\mathring{h}_{ab}$. Naturally, we can introduce the fiducial su(2)-valued triad $\mathring{e}^a_i$ and its co-triad $\mathring{\omega}_a^i$ which satisfy  $\mathring{h}_{ab}=\mathring{\omega}_a^i\mathring{\omega}_b^j\delta_{ij}$, $\mathring{h}^{ab}=\mathring{e}^a_i\mathring{e}^b_j\delta^{ij}$.

The classical background Hamiltonian is given by the classical Hamiltonian constraint smeared on some elementary cell $\mathcal{V}$ with the lapse function $N$,
\ba
\textbf{C}_{c}\big[N\big]=\int_{\mathcal{V}}d^3xN\mathcal{C}.
\ea
It is convenient to introduce the variables $c$, $p$ and $\pi$  by
\ba
  A_a^i=c\mathcal{V}^{-\frac{1}{3}}_o\mathring{\omega}_a^i,\quad E^a_i=p\det(\mathring{\omega})\mathcal{V}^{-\frac{2}{3}}_o\mathring{e}^a_i,\quad
  \tilde{\pi}=\pi\det(\mathring{\omega})\mathcal{V}^{-1}_o.\label{AEFRW}
  \ea
  where $\mathcal{V}_o\equiv\int_{\mathcal{V}}d^3x\det\big(\mathring{\omega}\big)$ is the volume of the elementary cell $\mathcal{V}$ measured by the fiducial metric $\mathring{h}_{ab}$.
It is easy to show that $c$, $p$ and $\pi$ are independent of the choice of the fiducial metric. Comparing (\ref{AEFRW}) with (\ref{FRWmetric}), we find $|p|\propto a^2$.  With these variables, the classical background Hamiltonian becomes
   \ba
  \textbf{C}_{c}[N]&=&N\Bigg[-\frac{1}{\kappa\gamma^2}\frac{3\sqrt{|p|}c^2}{F(\phi)}
   +\frac{\Big(\frac{3}{\kappa\gamma}F'(\phi)cp+F(\phi)\pi\Big)^2}{2|p|^{\frac{3}{2}}F(\phi)G(\phi)}
  +|p|^{\frac{3}{2}}V(\phi)\Bigg],\label{Hconstraint}
  \ea
  in which the conjugate variables satisfy
  \ba
  \{c,p\}=\frac{\kappa\gamma}{3},\qquad \{\phi,\pi\}=1.
  \ea
 It is not difficult to obtain the classical Friedmann equation, Klein-Gordon equation and Raychadhuri equation from the Hamiltonian constraint (\ref{Hconstraint}),
 \ba
&&\left(H+\frac{1}{2}\frac{\dot{F}(\phi)}{F(\phi)}\right)^2=\frac{\kappa}{3}\frac{\rho}{F^2(\phi)},\label{cFr}\\
&&\ddot{\phi}+3H\dot{\phi}+\frac{1}{2}\frac{\dot{G}(\phi)}{G(\phi)}\dot{\phi}
+\frac{1}{G(\phi)}\Big[F(\phi)V'(\phi)-2F'(\phi)V(\phi)\Big]=0,\label{cKG}\\
&&\dot{H}=\frac{1}{2F(\phi)}\Big[H\dot{F}(\phi)-\ddot{F}(\phi)-\kappa K(\phi)\dot{\phi}^2\Big],\label{cdotH}
\ea
 in which we set $N=1$ and the overdot denotes the differentiation with respect to the proper time, i.e.,$\dot{F}(\phi)\equiv\frac{dF(\phi)}{dt}$. $H$ is the Hubble parameter defined by $H:=\frac{\dot{p}}{2p}$, and $\rho$ is the effective energy density defined by
 \ba
 \rho:=\frac{1}{2}G(\phi)\dot{\phi}^2+F(\phi)V(\phi).\label{rhoe}
 \ea

\section{Quantisation ambiguities and effective theory of STT in the Jordan frame}\label{sec3}
Since in LQC there are no local operators corresponding to the curvature and the connection variable, we must regularise them as functions of holonomies which are well defined operators on the kinematic Hilbert space. In this section, we will quantise the Hamiltonian constraint using different regularisation prescriptions and derive the corresponding effective theory.

\subsection{Effective theory of the STT-curvature quantisation}
In the Hamiltonian constraint (\ref{newconstraint}), we denote the Euclidean term $\mathcal{C}_E$ and the Lorentzian term $\mathcal{C}_L$ as
\ba
\mathcal{C}_E&\equiv&\frac{F(\phi)}{2\kappa\sqrt{|\det E}|}\epsilon^{ij}_{~~k}E^a_iE^b_jF^k_{ab},\label{Eucterm}\\
\mathcal{C}_L&\equiv&\frac{F(\phi)}{\kappa\sqrt{|\det E|}}\Bigg(\gamma^2+\frac{1}{F^2(\phi)}\Bigg)
E^a_iE^b_jK^i_{[a}K^j_{b]}.
\ea
In the $k=0$ case, the spin connection $\Gamma_a^i$ vanishes, we have
$A_a^i=\gamma\!K_{a}^i$, hence the extrinsic curvature term in $\mathcal{C}_L$ becomes a multiple of the curvature term in $\mathcal{C}_E$,
\ba
2\gamma^2E^a_iE^b_jK^i_{[a}K^j_{b]}
=\epsilon^{ij}_{~~k}E^a_iE^b_jF_{ab}^k. \label{reEL0}
\ea
 For simplicity, the mainstream of LQC community prefer to quantise Lorentzian term in the same way as the  Euclidean term. The starting point is to rewrite the Hamiltonian in the $k=0$ case as
\ba
\textbf{C}^{(1)}&=&-\frac{\mathcal{V}^{\frac{2}{3}}_o\epsilon^{ij}_{~~k}
\mathring{e}^a_i\mathring{e}^b_j|p|^{2}F^k_{ab}}{2\kappa\gamma^2F(\phi)}
+\frac{\Big(\frac{1}{\kappa\gamma}F'(\phi)\mathcal{V}^{\frac{1}{3}}_o\mathring{e}^a_ipA_a^i+ F(\phi)\pi\Big)^2}{2F(\phi)G(\phi)}+|p|^{3}V(\phi),\nn\\\label{Hconstraintm}
\ea
where we have set $N=|p|^{\frac{3}{2}}$ for convenience of quantisation.
 The curvature $F_{ab}^{k}$ is regularised by calculating a closed holonomy around a plaquette  \cite{Ashtekar:2006b},
\ba
F_{ab}^k=
-2\text{Tr}\left(\frac{h^{(\bar{\mu})}_{\Box_{ij}}-\mathbb{I}}{\bar{\mu}^2}\tau^k\right)
\mathcal{V}_o^{-\frac{2}{3}}\mathring{\omega}_a^i\mathring{\omega}_b^j,\label{Fabk}
\ea
in which $\bar{\mu}\equiv\sqrt{\frac{\Delta}{p}}$ denotes the coordinate length of the edge of the plaquette with $\Delta\equiv4\sqrt{3}\pi\gamma G\hbar$ denoting the minimum nonzero area in loop quantum gravity, $\tau_k\equiv-\frac{i}{2}\sigma_k$
and $h^{(\bar{\mu})}_{\Box_{ij}}\equiv h^{(\bar{\mu})^{-1}}_{j}h^{(\bar{\mu})^{-1}}_{i}h^{(\bar{\mu})}_{j}h^{(\bar{\mu})}_{i}$ is the holonomy  around the plaquette $\Box_{ij}$.
Using the relation
\ba
 h^{(\bar{\mu})}_{i}=\cos\Big(\frac{\bar{\mu}c}{2}\Big)\mathbb{I}+2\sin\Big(\frac{\bar{\mu}c}{2}\Big)\tau_i,
\ea
the curvature can be expressed as
\ba
F_{ab}^k=\frac{\sin^2(\bar{\mu}c)}{\bar{\mu}^2}
\epsilon_{ij}^{~~k}\mathcal{V}_o^{-\frac{2}{3}}\mathring{\omega}_a^i\mathring{\omega}_b^j.\label{Fab}
\ea
In the $k=0$ case, the Ashtekar connection can be expressed in terms of $p$ and $F_{ab}^k$ via the following identity,
\ba
A_a^i
=-\frac{3}{4\kappa\gamma}\mathcal{V}^{\frac{1}{3}}_o\mathring{e}^b_j\epsilon^{ij}_{~~k}
\left\{p,F_{ab}^k\right\},\label{Aai}
\ea
then, using (\ref{Fab}), we obtain
\ba
A_a^i=\frac{\sin(2\bar{\mu}c)}{2\bar{\mu}}\mathcal{V}^{-\frac{1}{3}}_o\mathring{\omega}_a^i.\label{rAai}
\ea
Note that the identity in (\ref{Aai}) only holds for the $k=0$ FRW model, yet it can be generalised to other cosmological models and in the last section we will give an example by generalising (\ref{Aai}) to the $k=1$ model.

 For convenience of quantisation, we denote
\ba
b\equiv\bar{\mu}c,\quad v\equiv2\sqrt{3}~\text{sgn}(p)\bar{\mu}^{-3},
\ea
which satisfy $\{b,v\}=\frac{2}{\hbar}$, the variable $v$ is now proportional to the physical volume of the elementary cell.

The whole kinematical Hilbert space of STT $\mathcal{H}^{STT}_{kin}$ is a tensor product of the Hilbert space of the geometry and that of the scalar field, i.e., $\mathcal{H}^{STT}_{kin}=\mathcal{H}^{geo}_{kin}
\otimes\mathcal{H}^{\phi}_{kin}$. The kinematic quantum states are functions $\Psi(v,\phi)$ with finite norm
$\|\Psi\|^2\equiv\int\! d\phi\sum_{v}|\Psi(v,\phi)|^2$.
 We denote the orthonormal basis for $\mathcal{H}^{STT}_{kin}$ by $|v,\phi\rangle$ with
 \ba
 \langle\!v_1,\phi_1|v_2,\phi_2\rangle=
 \delta_{v_1,v_2}\delta(\phi_1,\phi_2).
 \ea
In $\mathcal{H}^{STT}_{kin}$, the operator $\hat{v}$ acts by simple multiplication and the operator $\widehat{\sin b}$ acts on the
basis $|v,\phi\rangle$ of the Hilbert space by \cite{Ashtekar:2006b},
\ba
\widehat
{\sin b}|v,\phi\rangle=\frac{1}{2i}\left[|v+2,\phi\rangle-|v-2,\phi\rangle\right].
\label{actsinb}
\ea
We choose the Schr\"{o}dinger representation for the scalar field, hence the operator
$\hat{\phi}$ acts on $|v,\phi\rangle$  by multiplication and its conjugate momentum $\hat{\pi}$ acts by differentiation. The quantum operator corresponding to the Hamiltonian
(\ref{Hconstraintm}) becomes
\ba
\hat{\textbf{C}}^{(1)}&=&-\frac{\Delta^2}{4\kappa\gamma^2}\widehat{\Bigg(\frac{1}{F(\phi)}\Bigg)}
\hat{v}(\widehat{\sin b})^2\hat{v}\nn\\
&&+\frac{3\Delta^2}{128\kappa^2\gamma^2}\widehat{\Bigg(\frac{(F'(\phi))^2}{F(\phi)G(\phi)}\Bigg)}
\Big[\widehat{\sin (2b)}\hat{v}+\hat{v}\widehat{\sin (2b)}\Big]^2\nn\\
&&+\frac{3\hbar}{32}
\Bigg[\widehat{\Bigg(\frac{F'(\phi)}{G(\phi)}\Bigg)}\hat{\pi}+\hat{\pi}\widehat{\Big(\frac{F'(\phi)}{G(\phi)}\Bigg)}\Bigg]
\Big[\widehat{\sin (2b)}\hat{v}+\hat{v}\widehat{\sin (2b)}\Big]  \nn\\
&&+\frac{1}{4}\Bigg[\hat{\pi}^2\widehat{\Bigg(\frac{F(\phi)}{G(\phi)}\Bigg)}
+\widehat{\Bigg(\frac{F(\phi)}{G(\phi)}\Bigg)}\hat{\pi}^2\Bigg]
+\frac{(\Delta)^{3}}{12}\hat{v}^2\hat{V}(\phi)\nn\\
&\equiv&\hat{C}^{(1)}_{1}+\hat{C}^{(1)}_{2}+\hat{C}^{(1)}_{3}+\hat{C}^{(1)}_{4}+\hat{C}^{(1)}_{5}.
\label{QHconstraintm}
\ea
The change of factor ordering in (\ref{QHconstraintm}) only affects trivial details but not the general properties of the effective dynamics.
To obtain well defined operators, we require that the functions  $\frac{1}{F(\phi)}$, $\frac{(F'(\phi))^2}{G(\phi)}$, $\frac{F'(\phi)}{G(\phi)}$, $\frac{F(\phi)}{G(\phi)}$ and $V(\phi)$ have no singularities. The operator $\hat{C}^{(1)}_1$ and $\hat{C}^{(1)}_2$ act on the basis by
\ba
\hat{C}^{(1)}_{1}|v,\phi\rangle&=&
\frac{\Delta^2}{16\kappa\gamma^2}\frac{v}{F(\phi)}
\Big[(v+4)|v+4,\phi\rangle-2v|v,\phi\rangle+(v-4)|v-4,\phi\rangle\Big],\nn\\\label{actC1}\\
\hat{C}^{(1)}_{2}|v,\phi\rangle&=&-\frac{3\Delta^2}{128\kappa\gamma^2}\frac{(F'(\phi))^2}{F(\phi)G(\phi)}
\Big[(v+2)(v+6)|v+8,\phi\rangle-2(v^2+4)|v,\phi\rangle\nn\\
&&\quad+(v-2)(v-6)|v-8,\phi\rangle\Big],\label{actC2}
\ea
where we have used the relation,
\ba
\widehat
{\sin (2b)}|v,\phi\rangle=\frac{1}{2i}\left[|v+4,\phi\rangle-|v-4,\phi\rangle\right].
\label{actsin2b}
\ea
Using equations (\ref{actsinb}) and (\ref{actsin2b}),
it is easy to write down the actions of the other operators on $|v,\phi\rangle$, which we will omit here for brevity.

In this subsection, to quantise the Hamiltonian constraint (\ref{Hconstraintm}), we first regularise the curvature, then, we use the regularised curvature in equation (\ref{Fab}) as well as the identity in (\ref{Aai}) to regularise the connection and obtain the result in (\ref{rAai}). After that, we promote the regularised variables to operators and obtain the quantum Hamiltonian operator (\ref{QHconstraintm}). Since in this quantisation prescription the regularisation of the connection relies on the regularised curvature and the Lorentzian term is quantised in the same way as the  Euclidean term,
 we call this quantisation prescription the STT-curvature quantisation.

Using the approach developed in \cite{Ashtekar:2010}, lengthy but straightforward calculation gives the effective Hamiltonian constraint,
\ba
\textbf{C}^{(1)}_{eff}&=&-\frac{\sqrt{3\Delta}}{2\kappa\gamma^2}\frac{1}{F(\phi)}|v|\sin^2 b
+\frac{\sqrt{3}}{\Delta^{\frac{3}{2}}}\frac{\Big(\frac{3\hbar}{8} F'(\phi)v\sin (2b)+ F(\phi)\pi\Big)^2}{F(\phi)G(\phi)|v|}\nn\\
&&+\frac{\Delta^{\frac{3}{2}}}{2\sqrt{3}}|v|V(\phi)
=0,\label{effHF}
\ea
in which we have reset $N=1$.
From (\ref{effHF}), we can obtain the canonical equations of motion,
\ba
\dot{b}&=&\text{sgn}(v)\frac{2}{\hbar}
\Bigg[-\frac{\sqrt{3\Delta}}{2\kappa\gamma^2}\frac{\sin^2 b}{F(\phi)}+\frac{9\sqrt{3}}{2^6\Delta^{\frac{3}{2}}}\frac{\Big(F'(\phi)\hbar \sin (2b)\Big)^2}{F(\phi)G(\phi)}-\frac{\sqrt{3}}{\Delta^{\frac{3}{2}}}\frac{F(\phi)}{G(\phi)}\frac{\pi^2}{|v|^2}\nn\\
&&\quad+\frac{\Delta^{\frac{3}{2}}}{2\sqrt{3}}V(\phi)\Bigg],\label{eomb}\\
\dot{v}&=&\text{sgn}(v)\frac{2}{\hbar}
\Bigg[
-\frac{3\sqrt{3}\hbar}{2\Delta^{\frac{3}{2}}}
\frac{\Big(\frac{3}{8}F'(\phi)\hbar v\sin (2b)+F(\phi)\pi\Big)}{F(\phi)G(\phi)}F'(\phi)\cos (2b) \nn\\
&&\quad+\frac{\sqrt{3\Delta}}{2\kappa\gamma^2}\frac{1}{F(\phi)}v\sin (2b)\Bigg],\label{eomv}\\
\dot{\phi}&=&\text{sgn}(v)\frac{3\sqrt{3}\hbar}{4\Delta^{\frac{3}{2}}}
\frac{F'(\phi)}{G(\phi)}\sin (2b)
+\frac{2\sqrt{3}}{\Delta^{\frac{3}{2}}}\frac{F(\phi)}{G(\phi)}\frac{\pi}{|v|},\label{eomphi}\\
\dot{\pi}&=&\frac{\sqrt{3\Delta}}{2\kappa\gamma^2}
\Bigg(\frac{1}{F(\phi)}\Bigg)'|v|\sin^2 b
-\frac{3\sqrt{3\Delta}}{16\kappa^2\gamma^2}\Bigg(\frac{(F'(\phi))^2}{F(\phi)G(\phi)}\Bigg)'
|v|\sin^2 (2b)\nn\\
&&-\text{sgn}(v)\frac{3\sqrt{3}\hbar}{4\Delta^{\frac{3}{2}}}
\Bigg(\frac{F'(\phi)}{G(\phi)}\Bigg)'\pi\sin (2b)-\frac{\sqrt{3}}{\Delta^{\frac{3}{2}}}\Bigg(\frac{F(\phi)}{G(\phi)}\Bigg)'\frac{\pi^2}{|v|}
-\frac{\Delta^{\frac{3}{2}}}{2\sqrt{3}}|v|V'(\phi).\nn\\\label{eompi}
\ea

Using these canonical equations of motion, we obtain the effective Friedmann equation and the effective Klein-Gordon equation,
\ba
&&\Bigg[H+\frac{1}{2}\frac{\dot{F}(\phi)}{F(\phi)}\Bigg(1-2\frac{\rho}{\rho_o}\Bigg)\Bigg]^2
=\frac{\kappa}{3}\frac{\rho}{F^2(\phi)}\Bigg(1-\frac{\rho}{\rho_o}\Bigg),\label{qFr1}\\
&&\ddot{\phi}+3H\dot{\phi}+\frac{1}{2}\frac{\dot{G}(\phi)}{G(\phi)}\dot{\phi}
+\frac{1}{G(\phi)}\Bigg[-2\Bigg(1-3\frac{\rho}{\rho_o}\Bigg)F'(\phi)V(\phi)+F(\phi)V'(\phi)\Bigg]\nn\\
&&-\frac{3F'(\phi)}{F(\phi)G(\phi)}\frac{\rho^2}{\rho_o}=0,\label{qKG1}
\ea
where $\rho_o\equiv \frac{3}{\Delta\kappa\gamma^2}$ and the definition of the effective energy density $\rho$ is given in equation (\ref{rhoe}).
From equations (\ref{qFr1}) and (\ref{qKG1}) we can get the effective Raychadhuri equation,
\ba
\dot{H}&=&\frac{1}{2}H\frac{\dot{F}(\phi)}{F(\phi)}\Bigg(1-6\frac{\rho}{\rho_o}\Bigg)
-\Bigg(\frac{\kappa}{2}\frac{K(\phi)}{F(\phi)}\dot{\phi}^2
+\frac{1}{2}\frac{\ddot{F}(\phi)}{F(\phi)}\Bigg)\Bigg(1-2\frac{\rho}{\rho_o}\Bigg)\nn\\
&&+\frac{\dot{F}(\phi)}{F(\phi)\rho_o}\Bigg[-3HG(\phi)\dot{\phi}^2
-\frac{3}{2}\frac{\dot{F}(\phi)}{F(\phi)}\rho\Bigg(1-4\frac{\rho}{\rho_o}\Bigg)
+3\dot{F}(\phi)V(\phi)\Bigg(1-2\frac{\rho}{\rho_o}\Bigg)\Bigg].\nn\\\label{qRe1}
\ea
Obviously,  the equations (\ref{qFr1}), (\ref{qKG1}) and (\ref{qRe1}) can reproduce the classical equations of motion when $\frac{\rho}{\rho_o}\rightarrow0$; moreover, these equations can also reproduce the effective equations of motion of the minimally coupled mainstream LQC when $F(\phi)=\!K(\phi)\equiv1$.

From the effective Friedmann equation (\ref{qFr1}), we learn that $\rho$ is upper bounded by $\rho_o$, which is similar as in the minimally coupled case, nevertheless, the Hubble parameter does not vanish at $\rho=\rho_o$, which is  unlike the minimally coupled case, it is easy to see that the necessary condition for the Hubble parameter to vanish is that the following requirement must be satisfied,
\ba
\frac{3}{4\kappa}\left(\dot{F}(\phi)\right)^2\Bigg(1-2\frac{\rho}{\rho_o}\Bigg)^2
=\rho\Bigg(1-\frac{\rho}{\rho_o}\Bigg).\label{ncm}
\ea
For the models in which $\rho\geq\frac{3}{4\kappa}(\dot{F}(\phi))^2$, equation (\ref{ncm}) can possibly be satisfied only when $\rho\geq\frac{3}{4}\rho_o$. Needless to say, whether the Hubble parameter can vanish during the effective evolution should be checked model by model in STT.

\subsection{Effective theory of the modified STT-curvature quantisation}\label{subsec3b}

 In this subsection, we quantise the Euclidean term and the Lorentzian term in different ways, mimicing the treatment adopted in the minimally coupled modified LQC in references \cite{Yang:2009,Assanioussi:2018} in which the Lorentzian term is regularised using the Thiemann's trick from the full theory \cite{Thiemann:1998,Thiemann:1996,Thiemann:2007}.

 The classical Hamiltonian constraint is written as
\ba
\textbf{C}^{(2)}&=&\frac{F(\phi)}{2\kappa}\mathcal{V}^{\frac{2}{3}}_o
  \mathring{e}^a_i\mathring{e}^b_j|p|^{2}\Bigg[\epsilon^{ij}_{~~k}F^k_{ab}
  -2\Bigg(1+\frac{1}{\gamma^2F^2(\phi)}\Bigg)A^i_{[a}A^j_{b]}\Bigg]\nn\\
   &&+\frac{\Big(\frac{1}{\kappa\gamma}F'(\phi)\mathcal{V}^{\frac{1}{3}}_o\mathring{e}^a_ipA_a^i+ F(\phi)\pi\Big)^2}{2F(\phi)G(\phi)}+|p|^{3}V(\phi)=0,\label{Hconstrainta}
\ea
where we have set $N=|p|^{\frac{3}{2}}$ for convenience of quantisation.

 The curvature $F_{ab}^k$ in the Hamiltonian constraint (\ref{Hconstrainta}) is regularised using equation (\ref{Fab}). Moreover,  since in the $k=0$ case the Thiemann's trick in equation (2.5) in reference \cite{Thiemann:1998} reduces to equation (\ref{Aai}), hence the regularised connection $A_a^i$ takes the same form as in equation (\ref{rAai}).

  The quantum operator corresponding to $\textbf{C}^{(2)}$ then becomes
\ba
\hat{\textbf{C}}^{(2)}&=&\frac{\Delta^2}{4\kappa}\widehat{F}(\phi)
\hat{v}(\widehat{\sin b})^2\hat{v}\nn\\
&&-\frac{\Delta^2}{64\kappa}
\Bigg[\frac{1+\gamma^2\hat{F}^2(\phi)}{\gamma^2\hat{F}(\phi)}
-\frac{3}{2\kappa}\widehat{\Bigg(\frac{(F'(\phi))^2}{F(\phi)G(\phi)}\Bigg)}\Bigg]
\Big[\widehat{\sin (2b)}\hat{v}+\hat{v}\widehat{\sin (2b)}\Big]^2\nn\\
&&+\frac{3\hbar}{32}
\Bigg[\widehat{\Bigg(\frac{F'(\phi)}{G(\phi)}\Bigg)}\hat{\pi}
+\hat{\pi}\widehat{\Bigg(\frac{F'(\phi)}{G(\phi)}\Bigg)}\Bigg]
\Big[\widehat{\sin (2b)}\hat{v}+\hat{v}\widehat{\sin (2b)}\Big]  \nn\\
&&+\frac{1}{4}\Bigg[\hat{\pi}^2\widehat{\Bigg(\frac{F(\phi)}{G(\phi)}\Bigg)}
+\widehat{\Bigg(\frac{F(\phi)}{G(\phi)}\Bigg)}\hat{\pi}^2\Bigg]
+\frac{(\Delta)^{3}}{12}\hat{v}^2\hat{V}(\phi)\nn\\
&\equiv&\hat{C}^{(2)}_{1}+\hat{C}^{(2)}_{2}+\hat{C}^{(2)}_{3}+\hat{C}^{(2)}_{4}+\hat{C}^{(2)}_{5}.\label{QHconstraintF}
\ea
 Using equations (\ref{actsinb}) and (\ref{actsin2b}), it is easy to write down how $\hat{\textbf{C}}^{(2)}$ acts on the kinematical Hilbert space and using these actions we can derive the effective Hamiltonian constraint,
\ba
\textbf{C}^{(2)}_{eff}&=&-\frac{\sqrt{3\Delta}}{2\kappa\gamma^2}\frac{1}{F(\phi)}|v|\sin^2 b\Big[1-\Big(1+\gamma^2F^2(\phi)\Big)\sin^2b\Big]\nn\\
&&+\frac{\sqrt{3}}{\Delta^{\frac{3}{2}}}\frac{\Big(\frac{3\hbar}{8} F'(\phi)v\sin (2b)+ F(\phi)\pi\Big)^2}{F(\phi)G(\phi)|v|}+\frac{\Delta^{\frac{3}{2}}}{2\sqrt{3}}|v|V(\phi)=0,\label{effH2}
\ea
where we have reset $N=1$. In the following, we call the quantisation prescription used in this subsection to quantise the Hamiltonian constraint (\ref{Hconstrainta}) the modified STT-curvature quantisation.

As can be expected, in the minimally coupled case $F(\phi)=K(\phi)\equiv1$, the effective Hamiltonian (\ref{effH2}) can reproduce the effective Hamiltonian constraint derived in \cite{Yang:2009,Assanioussi:2018}.  Moreover, if we set $F(\phi)=\sqrt{\kappa}\phi$ and $K(\phi)=\frac{\omega}{\sqrt{\kappa}\phi}$, the effective Hamiltonian constraint (\ref{effH2}) can reproduce the effective Hamiltonian constraint of Brans-Dicke theory in \cite{Song:2020} in which the Thiemann's trick is also used for regularisation.

The Hamilton's equations of motion can be directly obtained,
\ba
\dot{b}&=&\text{sgn}(v)\frac{2}{\hbar}
\Bigg[-\frac{\sqrt{3\Delta}}{2\kappa\gamma^2}\frac{1}{F(\phi)}\sin^2 b\Big[1-\Big(1+\gamma^2F^2(\phi)\Big)\sin^2 b\Big]\nn\\
&&+\frac{9\sqrt{3}}{64\Delta^{\frac{3}{2}}}\frac{\Big(F'(\phi)\hbar \sin (2b)\Big)^2}{F(\phi)G(\phi)}-\frac{\sqrt{3}}{\Delta^{\frac{3}{2}}}\frac{F(\phi)}{G(\phi)}\frac{\pi^2}{|v|^2}
+\frac{\Delta^{\frac{3}{2}}}{2\sqrt{3}}V(\phi)\Bigg],\label{eomb}\\
\dot{v}&=&\text{sgn}(v)\frac{2}{\hbar}
\Bigg[\frac{\sqrt{3\Delta}}{2\kappa\gamma^2}\frac{1}{F(\phi)}v\sin (2b)
\Big[1-2\sin^2b\Big(1+\gamma^2F^2(\phi)\Big)\Big]\nn\\
&&-\frac{3\sqrt{3}\hbar}{2\Delta^{\frac{3}{2}}}
\frac{\Big(\frac{3}{8}F'(\phi)\hbar v\sin (2b)+F(\phi)\pi\Big)}{F(\phi)G(\phi)}F'(\phi)\cos (2b) \Bigg],\label{eomv}\\
\dot{\phi}&=&\text{sgn}(v)\frac{3\sqrt{3}\hbar}{4\Delta^{\frac{3}{2}}}
\frac{F'(\phi)}{G(\phi)}\sin (2b)
+\frac{2\sqrt{3}}{\Delta^{\frac{3}{2}}}\frac{F(\phi)}{G(\phi)}\frac{\pi}{|v|},\label{eomphi}\\
\dot{\pi}&=&\frac{\sqrt{3\Delta}}{8\kappa\gamma^2}
\Bigg(\frac{K(\phi)}{G(\phi)}\Bigg)'|v|\sin^2 (2b)
-\frac{\sqrt{3\Delta}}{2\kappa}
\big(F(\phi)\big)'|v|\sin^4 b
\nn\\
&&-\text{sgn}(v)\frac{3\sqrt{3}\hbar}{4\Delta^{\frac{3}{2}}}
\Bigg(\frac{F'(\phi)}{G(\phi)}\Bigg)'\pi\sin (2b)-\frac{\sqrt{3}}{\Delta^{\frac{3}{2}}}\Bigg(\frac{F(\phi)}{G(\phi)}\Bigg)'\frac{\pi^2}{|v|}
-\frac{\Delta^{\frac{3}{2}}}{2\sqrt{3}}|v|V'(\phi).\nn\\\label{eompi}
\ea

To derive the effective Friedmann equation, we define a variable
\ba
\mathcal{A}:=1-2\big(1+\gamma^2F^2(\phi)\big)\sin^2 b,\label{defA}
\ea
from the definition of $\mathcal{A}$ in (\ref{defA}), we learn that $\mathcal{A}$ can take value in the range $(-\infty,1]$. Substituting equations (\ref{eomphi}) and (\ref{defA}) into the effective Hamiltonian constraint (\ref{effH2}), we find that (\ref{effH2}) can be rewritten as
\ba
\frac{\sqrt{3\Delta}|v|}{8\kappa\gamma^2}\frac{\big(\mathcal{A}-1\big)\big(\mathcal{A}+1\big)}{F(\phi)\big(1+\gamma^2F^2(\phi)\big)}
+\frac{\Delta^{\frac{3}{2}}}{2\sqrt{3}}\frac{|v|}{F(\phi)}\Bigg(\frac{1}{2}G(\phi)\dot{\phi}^2+F(\phi)V(\phi)\Bigg)=0,\nn\\\label{effH2n}
\ea
which can further reduce to
 \ba
\mathcal{A}^2=1-4\Big(1+\gamma^2F^2(\phi)\Big)\frac{\rho}{\rho_o}.\label{qconst2}
\ea
Since $\mathcal{A}^2\geq0$, we obtain $4\big(1+\gamma^2F^2(\phi)\big)\frac{\rho}{\rho_o}\leq1$, which means the function  $4\big(1+\gamma^2F^2(\phi)\big)\rho$ is upper bounded by $\rho_o$, hence the energy density $\rho$ is upper bounded by $\frac{\rho_o}{4}$.

Using equations (\ref{eomphi}) and (\ref{qconst2}), straightforward calculations give the effective Friedmann equation and the effective Klein-Gordon equation,
\ba
&&\left[H+\frac{1}{2}\frac{\dot{F}(\phi)}{F(\phi)}
\frac{\gamma^2F^2(\phi)+\mathcal{A}}{1+\gamma^2F^2(\phi)}\right]^2\nn\\
&=&\frac{\kappa}{3}\Bigg[\frac{\rho}{F^2(\phi)}
+\frac{\gamma^2\rho_o}{2\Big(1+\gamma^2F^2(\phi)\Big)}
\Bigg(\frac{1-\mathcal{A}}{1+\gamma^2F^2(\phi)}-\frac{2\rho}{\rho_o}\Bigg)
\Bigg]\mathcal{A}^2,\label{qFr2}\\
&&\ddot{\phi}+3H\dot{\phi}+\frac{1}{2}\frac{\dot{G}(\phi)}{G(\phi)}\dot{\phi}
+\frac{1}{G(\phi)}\Bigg[F(\phi)V'(\phi)+\frac{1-2\gamma^2F^2(\phi)-3\mathcal{A}}{1+\gamma^2F^2(\phi)}F'(\phi)V(\phi)\Bigg]\nn\\
&&+\frac{5\gamma^2\rho_oF(\phi)F'(\phi)}{4G(\phi)}
\Bigg(\frac{1-\mathcal{A}}{1+\gamma^2F^2(\phi)}\Bigg)^2
=0.\label{qKG2}
\ea
By taking time derivative of equation (\ref{qconst2}) and using equations (\ref{qFr2}) and (\ref{qKG2}), we derive the evolution equation of $\mathcal{A}$ and the effective Raychadhuri equation,
\ba
\mathcal{A}\dot{\mathcal{A}}&=&
\frac{1}{\rho_o}\Bigg[
6\Bigg(H+\frac{1}{2}\frac{\dot{F}(\phi)}{F(\phi)}\frac{\gamma^2F^2(\phi)+\mathcal{A}}{1+\gamma^2F^2(\phi)}\Bigg)
\Big(1+\gamma^2F^2(\phi)\Big)G(\phi)\dot{\phi}^2\nn\\
&&+\frac{5\rho_o\gamma^2F(\phi)\dot{F}(\phi)}{1+\gamma^2F^2(\phi)}(\mathcal{A}^2-\mathcal{A})-6\mathcal{A}\frac{\dot{F}(\phi)}{F(\phi)}\rho\Bigg],\label{dotA}\\
\mathcal{A}\dot{H}&=&-\frac{\big(\gamma^2F^2(\phi)\mathcal{A}+\mathcal{A}^2\big)}{2[1+\gamma^2F^2(\phi)]}
\frac{\ddot{F}(\phi)}{F(\phi)}
-\alpha\big(\phi,\dot{\phi}\big)
\Bigg[\frac{3}{4}\Bigg(\frac{\dot{F}(\phi)}{F(\phi)}\Bigg)^2+\frac{\kappa}{2}\frac{K(\phi)}{F(\phi)}\dot{\phi}^2\Bigg]\nn\\
&&+
\Bigg[\frac{3}{2}\beta\big(\phi,\dot{\phi}\big)
\Bigg(H+\frac{\dot{F}(\phi)}{2F(\phi)}
\frac{\gamma^2F^2(\phi)+\mathcal{A}}{1+\gamma^2F^2(\phi)}\Bigg)-\mathcal{A}H\Bigg]
\frac{\dot{F}(\phi)}{F(\phi)},\label{qRe2}
\ea
in which
\ba
\alpha\big(\phi,\dot{\phi}\big)&\equiv&\mathcal{A}-8\mathcal{A}\frac{\rho}{\rho_o}
-3\gamma^2F^2(\phi)\frac{\mathcal{A}-\mathcal{A}^2}{1+\gamma^2F^2(\phi)}\nn\\
&&+\frac{9}{4}\frac{\dot{F}(\phi)}{\kappa\rho_o}
\Bigg(H+\frac{\dot{F}(\phi)}{2F(\phi)}
\frac{\gamma^2F^2(\phi)+\mathcal{A}}{1+\gamma^2F^2(\phi)}\Bigg),\label{alpha}\\
\beta\big(\phi,\dot{\phi}\big)&\equiv&\Big(1+2\gamma^2F^2(\phi)\Big)\frac{\gamma^2F^2(\phi)+\mathcal{A}}{1+\gamma^2F^2(\phi)}-2\gamma^2F^2(\phi).
\ea
It is easy to see that these effective equations of motion can reproduce the classical equations of motion in the limit $\mathcal{A}\rightarrow1$.

In the following, we consider the models in which $\rho$ is lower bounded by zero. From equation (\ref{qconst2}), we find $\mathcal{A}$ can take value in the range $[-1,1]$ in these models.
Supposing that $\rho$ vanishes at $(\phi=\phi_v,\dot{\phi}=0)$, in the limit $\mathcal{A}\rightarrow-1$ we have
$H^2\rightarrow\frac{\kappa\gamma^2\rho_o}{3(1+\gamma^2F^2(\phi_v))^2}$, which corresponds to a contracting or an expanding de Sitter universe with an extremely large cosmological constant proportional to $\rho_o$.
Now, we have observed the role of the sign of $\mathcal{A}$ is a little bit subtle, on one hand, the sign of $\mathcal{A}$ cannot be directly observed, on the other hand, the sign of $\mathcal{A}$ makes physical sense.

In the effective Friedmann equation (\ref{qFr2}), we define
\ba
R(\phi,\dot{\phi},\mathcal{A}):=\frac{\rho}{F^2(\phi)}
+\frac{\gamma^2\rho_o}{2\big(1+\gamma^2F^2(\phi)\big)}
\Bigg(\frac{1-\mathcal{A}}{1+\gamma^2F^2(\phi)}-\frac{2\rho}{\rho_o}\Bigg).
\ea
Obviously, during the evolution we have $R(\phi,\dot{\phi},\mathcal{A})>0$, thus,
the Hubble parameter is expressed by
\ba
H_{1,2}=\pm\sqrt{\frac{\kappa}{3}}\sqrt{R(\phi,\dot{\phi},\mathcal{A})}\mathcal{A}
-\frac{1}{2}\frac{\dot{F}(\phi)}{F(\phi)}
\frac{\gamma^2F^2(\phi)+\mathcal{A}}{1+\gamma^2F^2(\phi)},\label{H12a}
\ea
in which $H_1$ corresponds to the plus sign and $H_2$ corresponds to the minus sign. Substituting $H_{1}$ or $H_{2}$ into the equations (\ref{dotA}) and (\ref{qRe2}), and imposing suitable initial conditions for $\phi$, $\dot{\phi}$ and $\mathcal{A}$, we can obtain two physically different solutions  $H_{1}(t)$ and $H_{2}(t)$. The initial conditions can be imposed  as follows. In the low energy limit, we impose $\phi_i$, $\dot{\phi}_i$ and $\mathcal{A}_{i}\rightarrow1$ for $H_1$ and the same values $\phi_i$, $\dot{\phi}_i$ but $\mathcal{A}_{i}\rightarrow-1$ for $H_2$, then, as discussed in the above paragraph, we obtain two different solutions, in which $H_{1}(t)$ describes an expanding classical universe in the low energy limit with $\mathcal{A}(t)\rightarrow1$ and a contracting de Sitter universe in the low energy limit with $\mathcal{A}(t)\rightarrow-1$, while $H_{2}(t)$ describes a contracting classical universe in the low energy limit with $\mathcal{A}(t)\rightarrow1$ and an expanding de Sitter universe in the low energy limit with $\mathcal{A}(t)\rightarrow-1$. It is obvious that neither of the solutions can describe a classical universe in both the contracting phase and the expanding phase of the universe. Note that the existence of two branches of solutions for the Hubble parameter still holds in the minimally coupled case $F(\phi)=\!K(\phi)\equiv1$, which means it is an inherent feature of the modified quantisation prescription, irrespective of the specific forms of the coupling functions.

\subsection{Effective theory of the STT-connection quantisation}\label{subsec3c}
In the above subsections, we regularise the connection variable by using the regularised curvature in (\ref{Fab}) and the identity (\ref{Aai}). In this subsection, we follow an different route by first regularising the connection variable and then use the result to regularise the curvature, in this case, the connection is expressed as functions of its holonomy along a open segment with coordinate length $2\bar{\mu}$,
\ba
A_a^i=-2\text{Tr}\left(\frac{h^{(2\bar{\mu})}_{j}-h^{(2\bar{\mu})^{-1}}_{j}}{4\bar{\mu}}\tau^j\right)
\mathcal{V}^{-\frac{1}{3}}_o\mathring{\omega}_a^i
=\frac{\sin (\bar{\mu}c)}{\bar{\mu}}\mathcal{V}^{-\frac{1}{3}}_o\mathring{\omega}_a^i,\label{opA}
\ea
in which the parameter $2\bar{\mu}$ is chosen only to agree with the minimally coupled mainstream LQC. It should be pointed out that other choices of length are also allowed. Now, the curvature is expressed by
\ba
F^i_{ab}=\epsilon^{i}_{~jk}A^j_aA^k_b=\frac{\sin^2(\bar{\mu}c)}{\bar{\mu}^2}
\epsilon^{i}_{~jk}\mathcal{V}_o^{-\frac{2}{3}}\mathring{\omega}_a^j\mathring{\omega}_b^k.\label{Fabo}
\ea
Since the regularisation of the curvature now depends on the regularised connection, in the following we call the quantisation prescription in this subsection the STT-connection quantisation.
Using (\ref{opA}) and (\ref{Fabo}), it is easy to derive the operator corresponding to the Hamiltonian (\ref{Hconstraintm}),
\ba
\hat{\textbf{C}}^{(3)}&=&-\frac{\Delta^2}{4\kappa\gamma^2}\widehat{\Bigg(\frac{1}{F(\phi)}\Bigg)}
\hat{v}(\widehat{\sin b})^2\hat{v}\nn\\
&&+\frac{3\Delta^2}{32\kappa^2\gamma^2}\widehat{\Bigg(\frac{F'(\phi)^2}{F(\phi)G(\phi)}\Bigg)}
\left[\hat{v}(\widehat{\sin b})+(\widehat{\sin b})\hat{v}\right]^2\nn\\
&&+\frac{3\hbar}{16}
\left[\widehat{\left(\frac{F'(\phi)}{G(\phi)}\right)}\hat{\pi}+\hat{\pi}\widehat{\left(\frac{F'(\phi)}{G(\phi)}\right)}\right]
\left[\widehat{\sin b}\hat{v}+\hat{v}\widehat{\sin b}\right]  \nn\\
&&+\frac{1}{4}\left[\hat{\pi}^2\widehat{\left(\frac{F(\phi)}{G(\phi)}\right)}
+\widehat{\left(\frac{F(\phi)}{G(\phi)}\right)}\hat{\pi}^2\right]
+\frac{(\Delta)^{3}}{12}\hat{v}^2\hat{V}(\phi)\nn\\
&\equiv&\hat{C}^{(3)}_{1}+\hat{C}^{(3)}_{2}+\hat{C}^{(3)}_{3}+\hat{C}^{(3)}_{4}+\hat{C}^{(3)}_{5}.\label{QHconstrainto}
\ea
Moreover, the operator corresponding to the Hamiltonian (\ref{Hconstrainta}) reads
\ba
\hat{\textbf{C}'}^{(3)}&=&\frac{\Delta^2}{4\kappa}\widehat{F}(\phi)
\hat{v}(\widehat{\sin b})^2\hat{v}\nn\\
&&-\frac{\Delta^2}{16\kappa}
\Bigg[\frac{1+\gamma^2\hat{F}^2(\phi)}{\gamma^2\hat{F}(\phi)}
-\frac{3}{2\kappa}\widehat{\left(\frac{(F'(\phi))^2}{F(\phi)G(\phi)}\right)}\Bigg]\bigg[\widehat{\sin (b)}\hat{v}+\hat{v}\widehat{\sin (b)}\bigg]^2\nn\\
&&+\hat{C}^{(3)}_{3}+\hat{C}^{(3)}_{4}+\hat{C}^{(3)}_{5}.\label{QHconstrainto2}
\ea
The operator $\hat{\textbf{C}}^{(3)}$ differs from the operator $\hat{\textbf{C}'}^{(3)}$ only in the factor orderings, which does not affect the effective description of the dynamics. The effective Hamiltonian is given by
\ba
\textbf{C}^{(3)}_{eff}&=&-\frac{\sqrt{3\Delta}}{2\kappa\gamma^2}\frac{1}{F(\phi)}|v|\sin^2 b
+\frac{\sqrt{3}}{\Delta^{\frac{3}{2}}}\frac{\big(\frac{3\hbar}{4} F'(\phi)v\sin b+ F(\phi)\pi\big)^2}{F(\phi)G(\phi)|v|}+\frac{\Delta^{\frac{3}{2}}}{2\sqrt{3}}|v|V(\phi)\nn\\
&=&0,\label{effHt}
\ea
where we have reset $N=1$.

The effective Friedmann equation, Klein-Gordon equation and Raychadhuri equation are given by \cite{Han:2019},
\ba
&&\Bigg(H+\frac{1}{2}\frac{\dot{F}(\phi)}{F(\phi)}\cos b\Bigg)^2
=\frac{\kappa}{3}\frac{\rho}{F^2(\phi)}\cos^2b,\label{qFr3}\\
&&\ddot{\phi}+3H\dot{\phi}+\frac{1}{2}\frac{\dot{G}(\phi)}{G(\phi)}\dot{\phi}
+\frac{1}{G(\phi)}\Big[(1-3\cos b)F'(\phi)V(\phi)+F(\phi)V'(\phi)\Big]=0,\nn\\\label{qKG3}\\
&&\dot{H}\cos b=
H\Bigg[\frac{3}{2}\Bigg(1-\frac{3\rho}{\rho_o}+\frac{F(\phi)V(\phi)}{\rho_o}\Bigg)-\cos b\Bigg]\frac{\dot{F}(\phi)}{F(\phi)}\nn\\
&&-\frac{\cos b}{2F(\phi)}\Bigg[\kappa K(\phi)\dot{\phi}^2\Bigg(1-\frac{2\rho}{\rho_o}\Bigg)
+\ddot{F}(\phi)\cos b-\frac{3}{2}\frac{\big(\dot{F}(\phi)\big)^2V(\phi)}{\rho_o}\Bigg],\label{qRe3}
\ea
in which the variable $\cos b$ is subject to the constraint,
\ba
&&\cos^2b=1-\frac{\rho}{\rho_o}. \label{qconst3}
\ea

Since $\cos b$ takes value in the range $[-1,1]$, from the constraint (\ref{qconst3}) we learn that $\rho$ is constrained to be nonnegative during the evolution. It is clear to see from equations (\ref{qFr3})$\sim$(\ref{qRe3}) that in the non-minimally coupled case with $F'(\phi)\neq0$ the classical equations of motion can be recovered in the low-energy limit with $\cos b\rightarrow 1$ but cannot be recovered in the low-energy limit with $\cos b\rightarrow -1$, while in the minimally coupled case with $F'(\phi)\equiv0$ the classical equations can be recovered in the low-energy limits with $\cos b\rightarrow \pm1$. In other words, in  the non-minimally coupled case different signs of $\cos b$ lead to different physics while in the minimally coupled case different signs of $\cos b$ do not cause any physical differences in the equations of motion.
It should be mentioned that this crucial fact was not noticed in previous literature such as \cite{Zhang:2013,Artymowski:2013,Jin:2018,Sharma:2019}, which caused the effective equations derived in these references not completely correct.

Using the equations of motion, straightforward derivation gives the evolution equation of $\cos b$,
\ba
\cos b\dot{\cos b}&=&\frac{3}{2\rho_o}\Bigg(HG(\phi)\dot{\phi}^2-\dot{F}(\phi)V(\phi)\cos b\Bigg).\label{eomcosb}
\ea

Notice that different choices of the length of the segment only affects the value of $\rho_o$ but does not affect any other term in the above equations, if we replace $2\bar{\mu}$ with $\delta\bar{\mu}$ in
equation (\ref{opA}), $\rho_o$ will be replaced by $\frac{4}{\delta^2}\rho_o$ in the effective equations.

Using the Friedmann equation (\ref{qFr3}), the Hubble parameter can be expressed as
\ba
H_{1,2}=\pm\sqrt{\frac{\kappa}{3}}\sqrt{\rho}\cos b
-\frac{1}{2}\frac{\dot{F}(\phi)}{F(\phi)}\cos b,\label{H12o}
\ea
in which $H_1$ corresponds to the plus sign and $H_2$ corresponds to the minus sign. It is easy to find that the sign of the Hubble parameter is the same as the sign of $\cos b$ in the $H_1$ branch and the opposite of the sign of $\cos b$ in the $H_2$ branch. Substituting $H_1$ or $H_2$ into equations (\ref{qKG3}), (\ref{qRe3}) and (\ref{eomcosb}), and specifying suitable initial values for $\phi$, $\dot{\phi}$ and $\cos b$,  we can obtain two physically different solutions  $H_{1}(t)$ and $H_{2}(t)$ in the non-minimally coupled case. The initial conditions can be imposed as follows. In the low-energy limit, we impose the same initial values of $\phi_i$ and $\dot{\phi}_i$ but opposite signs of $\cos b_i$ for $H_1$ and $H_2$ ($\cos b_i\rightarrow1$ for $H_1$ and $\cos b_i\rightarrow-1$ for $H_2$). From the equations of motion, we know that $H_{1}(t)$ and $H_{2}(t)$ must be different in the non-minimally coupled case because
they describe different expanding universes in the low-energy limit; moreover, in the limit $\cos b\rightarrow1$,  $H_{1}(t)$ describes an expanding classical universe  while $H_{2}(t)$ describes a contracting classical universe,
and neither of the solutions can describe a classical universe in both the contracting phase and the expanding phase.

We have noticed that there are some similarities between the effective equations of motion followed from the STT-connection quantisation and the ones followed from the modified STT-curvature quantisation. Nevertheless, there are some key differences between them. First, the existence of two physically different solutions of $H(t)$ under suitable initial conditions holds for the minimally coupled case  in the effective theory from the modified STT-curvature quantisation, while it no longer holds for the minimally coupled case in the effective theory from the STT-connection quantisation because as discussed above when $F'(\phi)\equiv0$ the two solutions of $H(t)$ are physically equivalent for opposite signs of $\cos b$. Second, due to the constraint (\ref{qconst3}), in the effective theory from the STT-connection quantisation the energy density always remains positive during the evolution for arbitrary $F(\phi)$, $K(\phi)$ and $V(\phi)$, yet in the effective theory from the modified STT-curvature quantisation it is not necessarily this case. Third,
equations (\ref{qFr3}) and (\ref{qconst3}) imply that in the effective theory from the STT-connection quantisation the Hubble parameter vanishes when $\cos b=0$, or, in other words, $\rho=\rho_o$, but in the effective theory from the modified STT-curvature quantisation the Hubble parameter does not vanish at a fixed energy density.

\section{Effective dynamics of the non-minimally coupled model}\label{sec4}
 To explicitly display the features of the effective dynamics of STT in the Jordan frame, in this section we consider a non-minimally coupled model which has been widely studied in cosmology (see for instance \cite{Tsujikawa:2004,Okada:2010}),  the coupling functions in this model read
\ba
 F(\phi)=1+\xi\kappa\phi^2,\quad K(\phi)\equiv1,\quad  V(\phi)=\frac{\lambda}{4}\phi^4, \label{FKV}
 \ea
in which the dimensionless coupling parameters $\xi,\lambda>0$.

\subsection{Big bang singularity in the classical dynamics of the non-minimally coupled model}
In the Jordan frame, the classical Friedmann equation of the non-minimally coupled model reads
\ba
\Bigg(H+\frac{\xi\kappa\phi\dot{\phi}}{1+\xi\kappa\phi^2}\Bigg)^2
=\frac{\kappa}{3}\frac{\rho}{(1+\xi\kappa\phi^2)^2},\label{nmsFr}
\ea
where
\ba
\rho&=&\frac{1}{2}\Big[1+(6\xi^2+\xi)\kappa\phi^2\Big]\dot{\phi}^2
+\frac{\lambda}{4}\big(1+\xi\kappa\phi^2\big)\phi^4\nn\\
&\equiv&\!
\mathcal{K}\big(\phi,\dot{\phi}\big)+\mathcal{P}(\phi),\label{nmsrho}
\ea
in which $\mathcal{K}\big(\phi,\dot{\phi}\big)$ denotes the kinetic energy density and $\mathcal{P}(\phi)$ denotes the potential energy density.
Since $\rho>3\xi^2\kappa\phi^2\dot{\phi}^2$, from equation (\ref{nmsFr}) it is easy to find that the Hubble parameter never vanishes. In the following, we show that the big bang singularity at which the Hubble parameter diverges is unavoidable in the classical dynamics of this model.

We define the following variables,
\ba
\tilde{a}:=\!a\sqrt{1+\xi\kappa\phi^2}, \quad \tilde{t}:=\int\sqrt{1+\xi\kappa\phi^2}dt,
\quad \tilde{\phi}:=\int\frac{\sqrt{1+(6\xi^2+\xi)\kappa\phi^2}}{1+\xi\kappa\phi^2}d\phi.\nn\\
\ea
Using these variables, the classical Friedmann equation and Raychadhuri equation can be rewritten as
\ba
\tilde{H}^2&=&\frac{\kappa}{3}\tilde{\rho},\label{tildeH2}\\
\frac{d\tilde{H}}{d\tilde{t}}&=&-\frac{\kappa}{2}\Bigg(\frac{d\tilde{\phi}}{d\tilde{t}}\Bigg)^2,\label{deotH}
\ea
in which
\ba
\tilde{H}&\equiv&\frac{1}{\tilde{a}}\frac{d\tilde{a}}{d\tilde{t}}
=\frac{1}{\sqrt{1+\xi\kappa\phi^2}}\Bigg(H+\frac{\xi\kappa\phi\dot{\phi}}{1+\xi\kappa\phi^2}\Bigg),\label{tildeH} \\
\tilde{\rho}&\equiv&\frac{1+(6\xi^2+\xi)\kappa\phi^2}{2(1+\xi\kappa\phi^2)^3}\dot{\phi}^2
+\frac{\lambda\phi^4}{4(1+\xi\kappa\phi^2)^2}\equiv\!
\tilde{\mathcal{K}}\big(\phi,\dot{\phi}\big)+\tilde{\mathcal{P}}(\phi).
 \ea
 We learn from these equations of motion that both $\tilde{H}$ and $\tilde{\rho}$ are monotonically decreasing with respect to $\tilde{t}$.

 Since
$
\frac{\lambda\phi^4}{4(1+\xi\kappa\phi^2)^2}<\frac{\lambda}{4\xi^2\kappa^2},
$
in the remote past when  $\tilde{H}^2>\frac{\lambda}{6\xi^2\kappa}$,
 $\tilde{\rho}$ would be dominated by $\tilde{\mathcal{K}}$, in this case, we obtain
 \ba
 &&\tilde{H}(\tilde{t})\simeq\frac{1}{\tilde{t}-\tilde{t}_r+\frac{1}{\tilde{H}_r}},\label{tildeHt}\\
 &&\Big|\tilde{\phi}(\tilde{t})-\tilde{\phi}(\tilde{t}_r)\Big|\simeq
 \sqrt{\frac{6}{\kappa}}\Big|\ln\big[\tilde{H}_r(\tilde{t}-\tilde{t}_r)+1\big]\Big|,\label{dtphi}
  \ea
 where $\tilde{t}_r$ denotes the value of $\tilde{t}$ at which $\tilde{H}^2=\frac{\lambda}{6\xi^2\kappa}$ and $\tilde{H}_r\equiv \tilde{H}(\tilde{t}_r)$. From equations (\ref{tildeHt}) and (\ref{dtphi}), we learn that $\tilde{H}$ and $\tilde{\phi}$ would reach infinity in a finite time interval before $\tilde{t}_r$.

 Since
 $\frac{d\tilde{\phi}}{d\phi}>0$ and $\frac{d|\tilde{\phi}|}{d\tilde{t}}<0$, we get $\frac{d|\phi|}{d\tilde{t}}<0$, hence, $F(\phi)$ would monotonically decrease with respect to $\tilde{t}$. Using equation (\ref{tildeH}), we find $H(\tilde{t})>\tilde{H}(\tilde{t})$. Moreover, since $\frac{d\tilde{t}}{dt}=\sqrt{1+\xi\phi^2}>1$,  $H(t)$ would also reach infinity in a finite time interval of $t$ before $\tilde{t}_r$. Therefore, we conclude that the big bang singularity is unavoidable in the classical dynamics.

\begin{figure}[h]
\centering
\includegraphics[height=5.5cm,width=7.8cm]{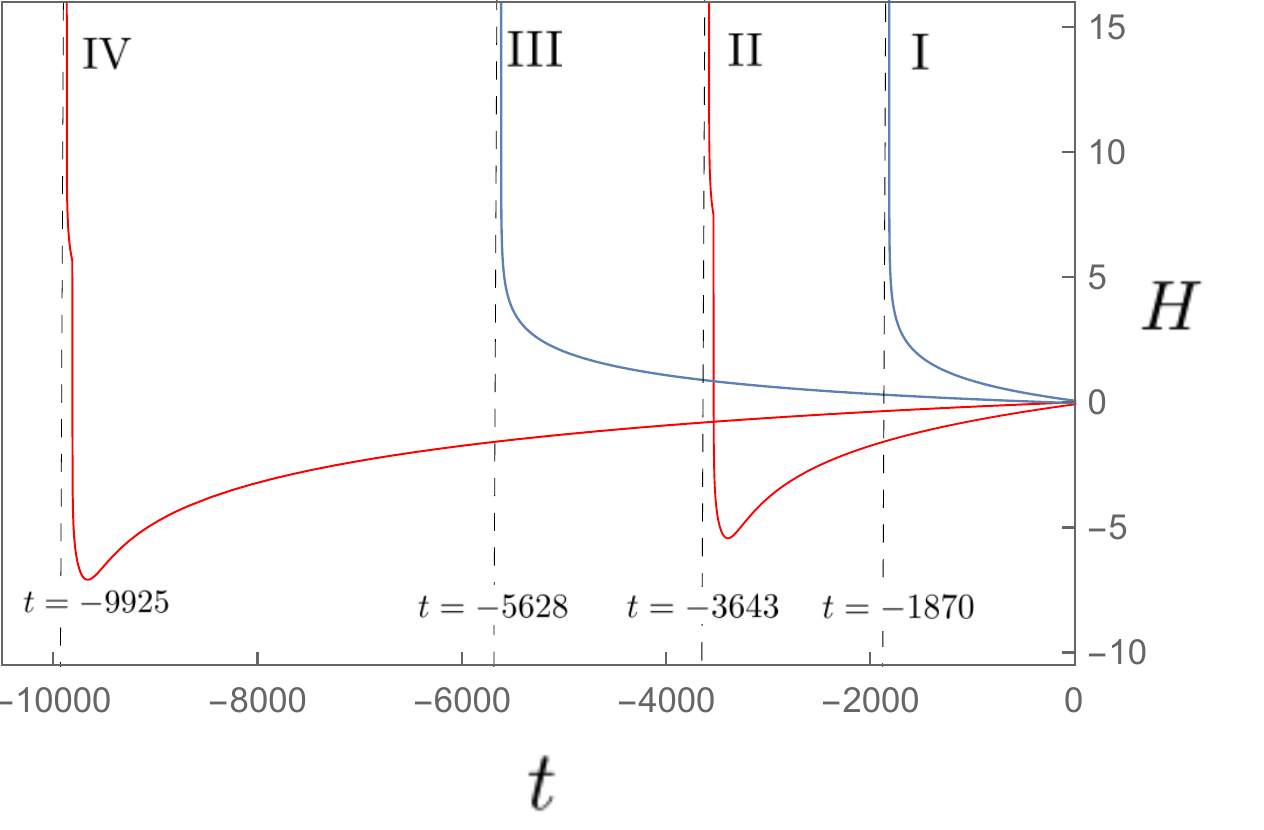}
\captionsetup{justification=raggedright}
\caption{The classical evolution of the Hubble parameter. The solid lines show the evolutions of the Hubble parameter before $t=0$. In the four curves, the initial values are chosen, respectively, as $\phi_i=0.05\kappa^{-\frac{1}{2}}$, $\dot{\phi}_i=-6.91\times10^{-6}\kappa$ (curve I), $\phi_i=0.05\kappa^{-\frac{1}{2}}$, $\dot{\phi}_i=6.91\times10^{-6}\kappa$ (curve II), $\phi_i=0.072\kappa^{-\frac{1}{2}}$, $\dot{\phi}_i=-2.59\times10^{-6}\kappa$ (curve III) and $\phi_i=0.05\kappa^{-\frac{1}{2}}$, $\dot{\phi}_i=2.59\times10^{-6}\kappa$ (curve IV).  The value of the initial energy density, $\rho_i=10^{-4}\kappa^2$, is the same for all curves. Curves I and II represent the initially kinetic energy dominated state while curves III and IV represent the initially potential energy dominated state. The dashed lines denote the moment at which the Hubble parameter reaches infinity. For clarity, the values of the quantities in this figure (and all the other figures and tables in the following subsections) are given by setting $\kappa=\hbar=c=1$.}
\label{fig1}
\end{figure}

 In Figure \ref{fig1}, we plot the classical evolution of the Hubble parameter before the initial moment.  We specify $\lambda=0.13$ and $\xi=1.6\times10^4$. The self-interacting potential with $\lambda=0.13$ is a good approximation of the Higgs potential when the scalar field is much larger than its mass, and
 the large value $\xi=1.6\times10^4$ is chosen for this model to generate predictions completely agreeing with the current observation of slow-roll inflation \cite{Felice:2011}.

 \clearpage

\subsection{Effective dynamics from the STT-curvature quantisation}
Substituting the functions in (\ref{FKV}) into equations (\ref{qFr1}), (\ref{qKG1}) and (\ref{qRe1}), we obtain
the effective Friedmann equation, Klein-Gordon equation, and Raychadhuri equation of the non-minimally coupled model in the STT-curvature quantisation,
\ba
&&\Bigg[H+\frac{\xi\kappa\phi\dot{\phi}}{1+\xi\kappa\phi^2}\Bigg(1-2\frac{\rho}{\rho_o}\Bigg)\Bigg]^2
=\frac{\kappa}{3}\frac{\rho}{(1+\xi\kappa\phi^2)^2}\Bigg(1-\frac{\rho}{\rho_o}\Bigg),\label{nmsFrm}\\
&&\ddot{\phi}+3H\dot{\phi}+\frac{(6\xi^2+\xi)\kappa\phi\dot{\phi}^2+\frac{3\rho}{\rho_o}\lambda\xi\kappa\phi^5
+\lambda\phi^3}{1+(6\xi^2+\xi)\kappa\phi^2}\nn\\
&&\quad-\frac{6\xi\kappa\phi\rho^2}{\rho_o(1+\xi\kappa\phi^2)\Big(1+(6\xi^2+\xi)\kappa\phi^2\Big)}=0,\label{nmsKGm}\\
&&\dot{H}=\frac{1}{1+\xi\kappa\phi^2}\Bigg[H\xi\kappa\phi\dot{\phi}\Bigg(1-6\frac{\rho}{\rho_o}\Bigg)
-\Bigg(\frac{\kappa}{2}\dot{\phi}^2
+\xi\kappa\Big(\dot{\phi}^2+\phi\ddot{\phi}\Big)\Bigg)\Bigg(1-2\frac{\rho}{\rho_o}\Bigg)\Bigg]\nn\\
&&\qquad+\frac{\xi\kappa\phi\dot{\phi}}{(1+\xi\kappa\phi^2)\rho_o}
\Bigg[-6H\Big(1+(6\xi^2+\xi)\kappa\phi^2\Big)\dot{\phi}^2
-\frac{6\xi\kappa\phi\dot{\phi}}{1+\xi\kappa\phi^2}\rho
\Bigg(1-4\frac{\rho}{\rho_o}\Bigg)\nn\\
&&\qquad+3\lambda\xi\kappa\phi^5\dot{\phi}\Bigg(1-2\frac{\rho}{\rho_o}\Bigg)\Bigg].
\label{nmsRem}
\ea

From equations (\ref{nmsrho}) and  (\ref{nmsFrm}), we find that
\ba
0\leq\rho\leq\rho_o,\quad
(6\xi^2+\xi)\kappa\phi^2\dot{\phi}^2<2\rho_o,
\quad
\lambda\big(1+\xi\kappa\phi^2\big)\phi^4\leq4\rho_o. \label{inequalm}
\ea
Substituting these inequalities into (\ref{nmsFrm}) and (\ref{nmsKGm}), we get
\ba
|H|&<&\Bigg(\frac{1}{2\sqrt{3}}+\frac{2}{3\sqrt{3}}C(\xi)\Bigg)\sqrt{\kappa\rho_o},\label{abHm}\\
\Big|\dot{H}\Big|&<&\Bigg[1+2\xi
+\Bigg(6\sqrt{2}+\frac{4}{\sqrt{3}}\Bigg)C(\xi)
+\Bigg(\frac{16}{3\sqrt{3}}+\frac{51}{2}\Bigg)C^2(\xi)\Bigg]\kappa\rho_o,\label{abdotHm}
\ea
where $C(\xi)\equiv\sqrt{\frac{\xi^2}{6\xi^2+\xi}}$. In the small coupling limit, we have
$\lim\limits_{\xi\rightarrow0}C(\xi)\rightarrow0$ and in the large coupling limit we have $\lim\limits_{\xi\rightarrow\infty}C(\xi)\rightarrow\sqrt{\frac{1}{6}}$,  equations (\ref{inequalm}), (\ref{abHm}) and (\ref{abdotHm}) imply that the effective energy density, the Hubble parameter and the spacetime curvature $R\equiv6\big(2H^2+\dot{H}\big)$ do not diverge for finite $\xi$, hence, the spacetime singularity is absent in the effective theory of the STT-curvature quantisation.

\begin{figure}[h]
\begin{center}
\subfloat[\label{fig2a}
]{\includegraphics[height=4cm]{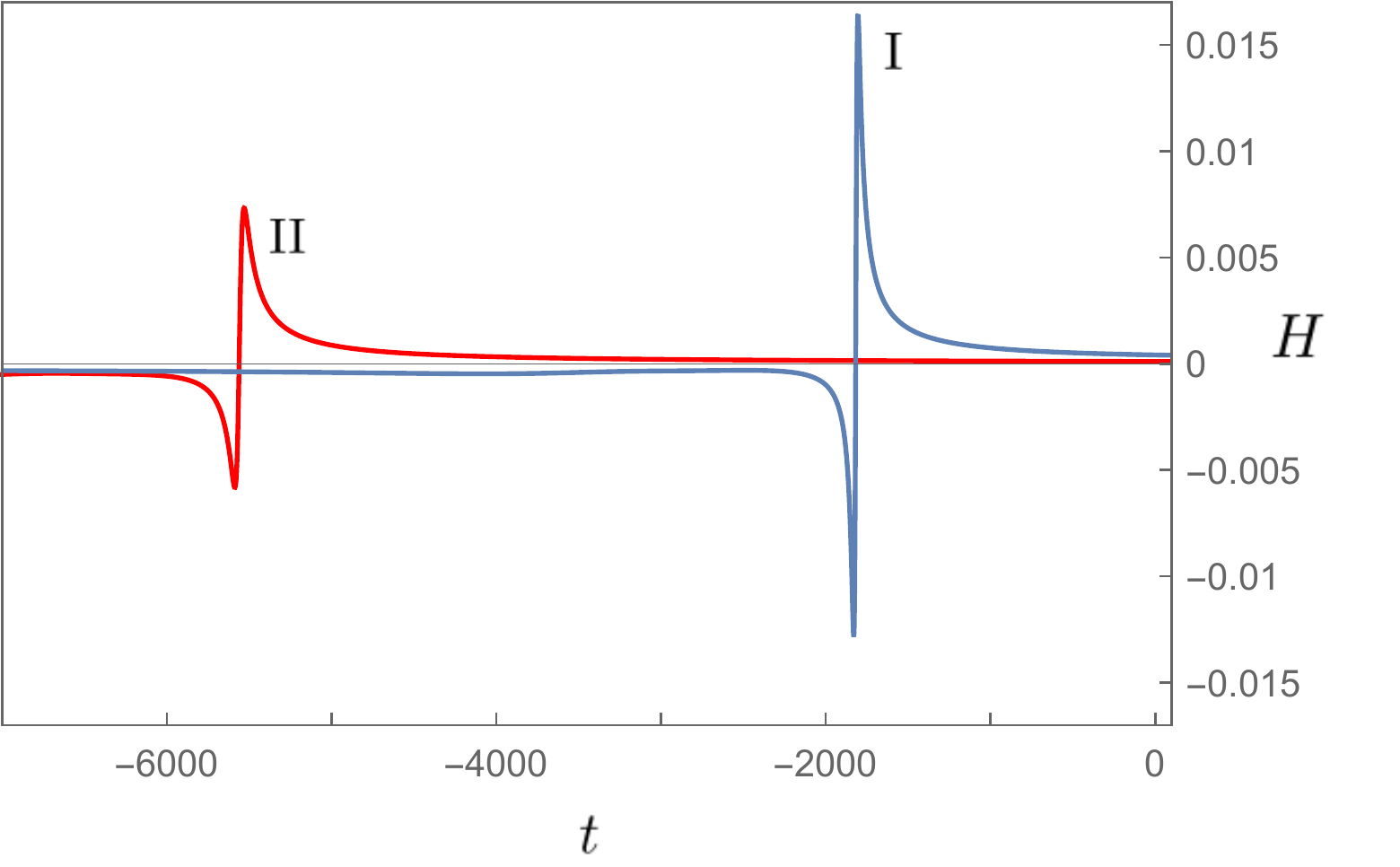}}
\hspace{1cm}
\subfloat[\label{fig2b}]{\includegraphics[height=4cm]{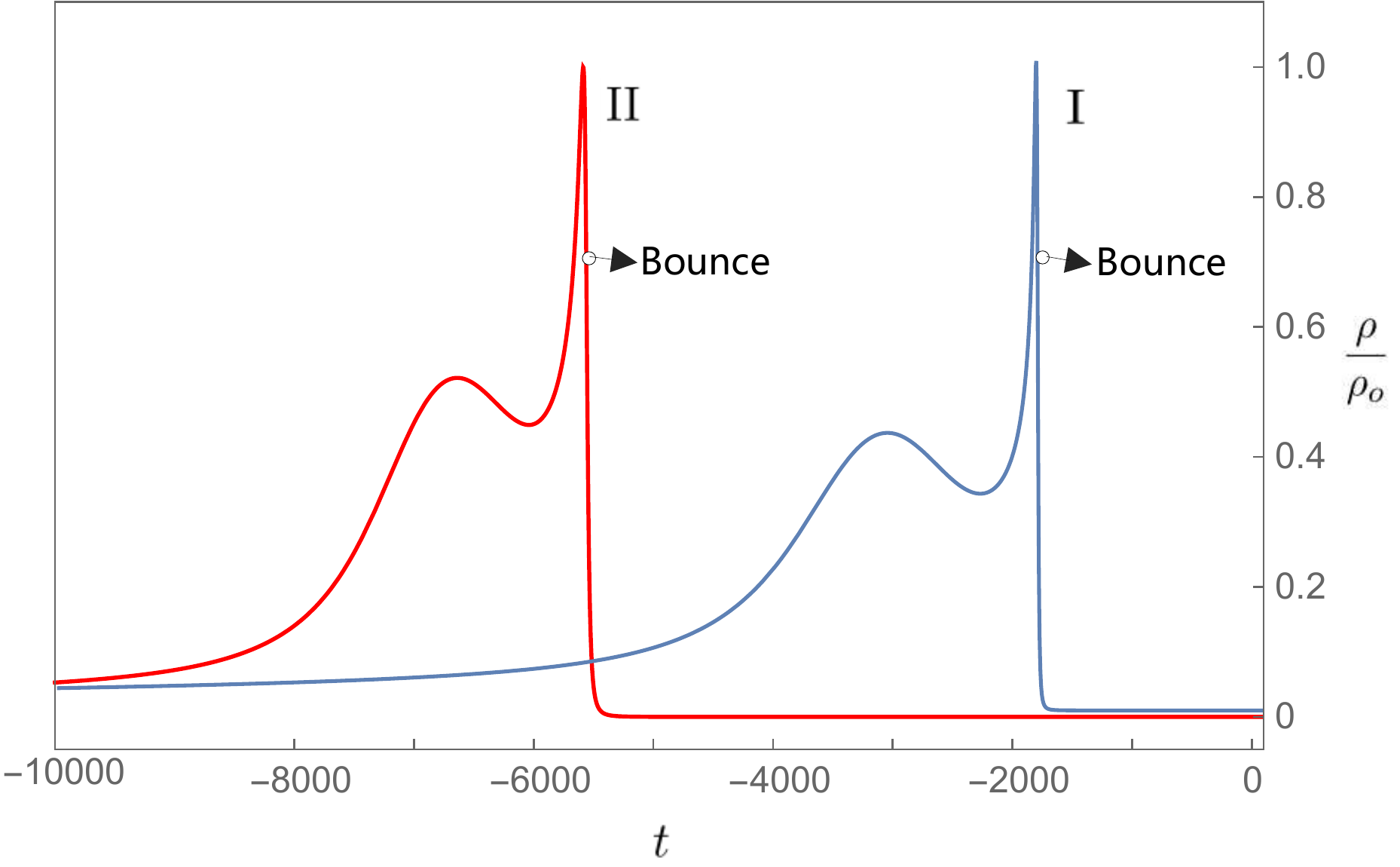}}
\captionsetup{justification=raggedright}
\caption{Figure \ref{fig2a} shows the evolution of the Hubble parameter in the effective theory of the STT-curvature quantisation for the non-minimally coupled model with $\lambda=0.13$, $\xi=1.6\times10^4$. The initial conditions are $\phi_i=0.05\kappa^{-\frac{1}{2}}$, $\dot{\phi}_i=-6.91\times10^{-6}\kappa$  for curve I and $\phi_i=0.072\kappa^{-\frac{1}{2}}$, $\dot{\phi}_i=-2.59\times10^{-6}\kappa$ for curve II, which represent the initially kinetic and potential energy dominated state respectively.  Figure \ref{fig2b} shows the corresponding evolution of $\frac{\rho}{\rho_o}$ for curve I and curve II, in which the hollow dot denotes event of the quantum bounce.}
\label{fig2}
\end{center}
\end{figure}

With the help of numerical methods, we find that in this model an expanding classical universe is born out of a contracting classical universe  under different initial conditions in the effective theory of the STT-curvature quantisation (see for instance Figure \ref{fig2}).

In Table \ref{table1}, we list the relations between different initial conditions and the values of some parameters at the bounce. Detailed numerical results reveal some interesting facts for almost all initial conditions (most of which are not listed in the table for brevity), for instance,  the energy density at the bounce is extremely kinetic energy density dominated, and the bounces take place at energy densities about $0.75\rho_o$, etc.. These facts can be semi-quantitatively explained as follows.

\begin{center}
\begin{table}[h]
\begin{tabular}{|c|c|c|c|c|c|}\hline
  $\theta_i$&$t_{b}$&$\rho_{b}/\rho_o$& $\phi_{b}$ & $\dot{\phi}_{b}$&$\mathcal{P}(\phi_b)/\rho_{b}$  \\
  \hline
  $0$&$-43.1$&$0.75+9.1\times10^{-7}$ & $-2.74\times10^{-2}$ & $1.84\times10^{-2}$&$1.22\times10^{-9}$ \\
  \hline
  $\pi/6$&$-5511.7$&$0.75+8.7\times10^{-7}$ & $-2.77\times10^{-2}$ & $1.81\times10^{-2}$&$1.31\times10^{-9}$ \\
  \hline
  $\pi/3$&$-10680.9$&$0.75+8.4\times10^{-7}$ & $-2.88\times10^{-2}$ & $1.74\times10^{-2}$ &$1.65\times10^{-9}$ \\
  \hline
  $\pi/2$&$-36053.2$&$0.75+8.2\times10^{-7}$ & $-3.08\times10^{-2}$ & $1.63\times10^{-2}$&$2.43\times10^{-9}$ \\
  \hline
  $2\pi/3$&$-5713.4$&$0.75+9.7\times10^{-5}$ & $2.75\times10^{-1}$ & $-1.82\times10^{-3}$&$1.16\times10^{-3}$ \\
  \hline
  $5\pi/6$&$-2804.4$&$0.75+2.7\times10^{-5}$ & $2.21\times10^{-1}$ & $-2.27\times10^{-3}$&$3.13\times10^{-4}$ \\\hline
  \end{tabular}
\captionsetup{justification=raggedright}
\caption{Relations between the initial conditions and values of parameters at the quantum bounce in the effective theory of the STT-curvature quantisation. In the table, we define
$\theta_i:=\! \arcsin\left(\text{sgn}(\dot{\phi}_i)\sqrt{\frac{\mathcal{P}(\phi_i)}{\rho_i}}\right)$ in which $\mathcal{P}(\phi_i)$ is the initial potential energy density and $\rho_i=10^{-4}\kappa^2$ is the initial energy density. We only consider the initial conditions with $\phi_i>0$ because of the symmetry of solutions of the equations of motion of this model.  The subscript $b$ means the values are evaluated at the instant of the quantum bounce. }\label{table1}
\end{table}
\end{center}

When $\rho\leq\rho_o$ and $H\geq0$, equation (\ref{nmsKGm}) can be written as
\ba
\dot{f}\big(\phi,\dot{\phi}\big)+g\big(\phi,\dot{\phi}\big)f\big(\phi,\dot{\phi}\big)
+h\big(\phi,\dot{\phi}\big)=0,\label{eomk}
\ea
in which
\ba
f\big(\phi,\dot{\phi}\big)&\equiv&\sqrt{1+(6\xi^2+\xi)\kappa\phi^2}\dot{\phi},\nn\\
g\big(\phi,\dot{\phi}\big)&\equiv&\frac{\sqrt{3\kappa}}{1+\xi\kappa\phi^2}\Bigg[
\sqrt{\rho}\sqrt{1-\frac{\rho}{\rho_o}}-\sqrt{3\kappa}\xi\phi\dot{\phi}\Bigg(1-\frac{\rho}{\rho_o}\Bigg)\Bigg],\nn\\
h\big(\phi,\dot{\phi}\big)&\equiv&\frac{\frac{3\rho}{2\rho_o}\lambda\xi\kappa\phi^5
+\lambda\phi^3}{\sqrt{1+(6\xi^2+\xi)\kappa\phi^2}}. \label{defnmsm}
\ea
In the case $\rho\ll\rho_o$, $g\big(\phi,\dot{\phi}\big)$ and $h\big(\phi,\dot{\phi}\big)$ can be approximated by
\ba
g\big(\phi,\dot{\phi}\big)\approx\frac{\sqrt{3\kappa}}{1+\xi\kappa\phi^2}\big(
\sqrt{\rho}-\sqrt{3\kappa}\xi\phi\dot{\phi}\big),\quad
h\big(\phi,\dot{\phi}\big)\approx\frac{\lambda\phi^3}{\sqrt{1+(6\xi^2+\xi)\kappa\phi^2}}.
\ea

In the large $\xi$ case, we have $\sqrt{1+(6\xi^2+\xi)\kappa\phi^2}\approx\!\sqrt{6\kappa}\xi|\phi|$ for $\phi\neq0$, hence
\ba
&&\Big|g\big(\phi,\dot{\phi}\big)f\big(\phi,\dot{\phi}\big)\Big|\approx\!
\frac{\sqrt{6\kappa}}{1+\xi\kappa\phi^2}\Big(\sqrt{\mathcal{K}+\mathcal{P}}
-\text{sgn}\big(\phi\dot{\phi}\big)\sqrt{\mathcal{K}}\Big)\sqrt{\mathcal{K}},\nn\\
&&\Big|h\big(\phi,\dot{\phi}\big)\Big|\approx\!\frac{\sqrt{\lambda}}{\sqrt{6}\xi}\frac{\sqrt{\mathcal{P}}}{\sqrt{\kappa}\sqrt{1+\xi\kappa\phi^2}},
\ea
in which the definition of the kinetic energy density $\mathcal{K}$ and the potential energy density $\mathcal{P}$ is given in (\ref{nmsrho}). It is not difficult to find that $|gf|>|h|$ when
\ba
\mathcal{K}>\frac{\alpha^2\sqrt{\mathcal{P}}}{\kappa^2\sqrt{\mathcal{P}}-2\alpha\kappa\text{sgn}\big(\phi\dot{\phi}\big)\sqrt{1+\xi\kappa\phi^2}},
\ea
in which $\alpha\equiv\frac{\sqrt{\lambda}}{6\sqrt{6}\xi}$. In particular, we have $|gf|>|h|$  for $\mathcal{K}>\frac{\alpha^2}{\kappa^2}$ and $\phi\dot{\phi}<0$, in this case, equation (\ref{eomk}) can be approximated by $\dot{f}+gf\approx0$, note that $2\mathcal{K}=f^2$, we obtain
\ba
\dot{\mathcal{K}}+\frac{2\sqrt{6\kappa}}{1+\xi\kappa\phi^2}\Big(\sqrt{\mathcal{K}+\mathcal{P}}
+\sqrt{\mathcal{K}}\Big)\mathcal{K}\approx0,
\ea
 hence, the kinetic energy density monotonically increases as the time reverses,
For $\lambda=0.13$, $\xi=1.6\times10^{4}$, we have $\alpha=1.53\times10^{-6}$ and $\frac{\alpha^2}{\kappa^2\rho_o}=9.05\times10^{-15}$, therefore,  the condition $\mathcal{K}>\frac{\alpha^2}{\kappa^2}$ can be easily satisfied during the evolution. As a result, the kinetic energy density becomes more and more dominant as the time reverses. Although the quantum gravity effects become important when $\rho$ approaches $\rho_o$, numerical analysis shows that the quantum gravity effects only play an important role in an extremely short time interval, hence the kinetic energy density still dominates the energy density at the bounce for almost all solutions.
From equation (\ref{nmsFrm}), we find that the Hubble parameter vanishes when
\ba
\sqrt{\rho}\sqrt{1-\frac{\rho}{\rho_o}}=\sqrt{3\kappa}\xi\left|\phi\dot{\phi}\Bigg(1-2\frac{\rho}{\rho_o}\Bigg)\right|.\label{eomHv}
\ea
In the kinetic energy dominated case, we have $\sqrt{\rho}\approx\mathcal{\sqrt{K}}\approx\sqrt{3\kappa}\xi|\phi\dot{\phi}|$, and equation ({\ref{eomHv}}) can be approximated by
\ba
\sqrt{1-\frac{\rho}{\rho_o}}\approx\left|1-2\frac{\rho}{\rho_o}\right|,
\ea
which leads to $\rho_b\approx0.75\rho_o$.

\subsection{Effective dynamics from the modified STT-curvature quantisation}

Substituting (\ref{FKV}) into equations (\ref{qconst2})$\sim$(\ref{qRe2}), we obtain the effective equations of motion from the modified STT-curvature quantisation of the non-minimally coupled model. The effective Friedmann equation and  Klein-Gordon equation are expressed as
\ba
&&\left[H+\frac{\xi\kappa\phi\dot{\phi}}{1+\xi\kappa\phi^2}
\frac{\mathcal{A}+\gamma^2(1+\xi\kappa\phi^2)^2}{1+\gamma^2(1+\xi\kappa\phi^2)^2}
\right]^2\nn\\
&&=\frac{\kappa}{3}\left[\frac{\rho}{(1+\xi\kappa\phi^2)^2}
+\frac{\gamma^2\rho_o}{2\big[1+\gamma^2(1+\xi\kappa\phi^2)^2\big]}
\Bigg(\frac{1-\mathcal{A}}{1+\gamma^2(1+\xi\kappa\phi^2)^2}-\frac{2\rho}{\rho_o}\Bigg)
\right]
\mathcal{A}^2,\nn\\\label{nmsFra}\\
&&\ddot{\phi}+3H\dot{\phi}
+\frac{1}{1+(6\xi^2+\xi)\kappa\phi^2}
\Bigg[(6\xi^2+\xi)\kappa\phi\dot{\phi}^2+\frac{3(1-\mathcal{A})\lambda\xi\kappa\phi^5}{2\big[1+\gamma^2(1+\xi\kappa\phi^2)^2\big]}
+\lambda\phi^3\Bigg]\nn\\
&&+\frac{5\gamma^2\rho_o(1+\xi\kappa\phi^2)\xi\kappa\phi}{2\big[1+(6\xi^2+\xi)\kappa\phi^2\big]}
\Bigg(\frac{1-\mathcal{A}}{1+\gamma^2(1+\xi\kappa\phi^2)^2}\Bigg)^2
=0.\label{nmsKGa}
\ea
where $\mathcal{A}$ is subject to the following constraint
\ba
\mathcal{A}^2=1-4\Big[1+\gamma^2(1+\xi\kappa\phi^2)^2\Big]\frac{\rho}{\rho_o}.\label{nmsconstA}
\ea
The evolution equation of $\mathcal{A}$ and the effective Raychadhuri equation read
\ba
\mathcal{A}\dot{\mathcal{A}}&=&\frac{1}{\rho_o}\Bigg[
3\Bigg(H+\frac{\xi\kappa\phi\dot{\phi}}{1+\xi\kappa\phi^2}
\frac{\mathcal{A}+\gamma^2(1+\xi\kappa\phi^2)^2}{1+\gamma^2(1+\xi\kappa\phi^2)^2}
\Bigg)\Big(1+\gamma^2(1+\xi\kappa\phi^2)^2\Big)\nn\\
&&\quad\times\Big(1+(6\xi^2+\xi)\kappa\phi^2\Big)\dot{\phi}^2
+\frac{5\gamma^2\rho_o(1+\xi\kappa\phi^2)\xi\kappa\phi\dot{\phi}}{1+\gamma^2(1+\xi\kappa\phi^2)^2}(\mathcal{A}^2-\mathcal{A})\nn\\
&&\quad
-12\frac{\xi\kappa\phi\dot{\phi}}{1+\xi\kappa\phi^2}\rho\mathcal{A}\Bigg],\label{nmsaeomA}\\
\mathcal{A}\dot{H}&=&-\frac{\gamma^2(1+\xi\kappa\phi^2)^2+\mathcal{A}}{1+\gamma^2(1+\xi\kappa\phi^2)^2}
\frac{\xi\kappa\mathcal{A}(\dot{\phi}^2+\phi\ddot{\phi})}{1+\xi\kappa\phi^2}\nn\\
&&-\alpha\big(\phi,\dot{\phi}\big)
\Bigg(3\frac{\xi^2\kappa^2\phi^2\dot{\phi}^2}{(1+\xi\kappa\phi^2)^2}+\frac{\kappa}{2}\frac{\dot{\phi}^2}{1+\xi\kappa\phi^2}\Bigg)\nn\\
&&+
\Bigg[\frac{3}{2}\beta\big(\phi,\dot{\phi}\big)
\Bigg(H+\frac{\xi\kappa\phi\dot{\phi}}{1+\xi\kappa\phi^2}
\frac{\mathcal{A}+\gamma^2(1+\xi\kappa\phi^2)^2}{1+\gamma^2(1+\xi\kappa\phi^2)^2}\Bigg)-\mathcal{A}H\Bigg]
\frac{2\xi\kappa\phi\dot{\phi}}{1+\xi\kappa\phi^2},\nn\\\label{nmsRea}
\ea
in which
\ba
\alpha\big(\phi,\dot{\phi}\big)&=&\mathcal{A}-8\mathcal{A}\frac{\rho}{\rho_o}
+\frac{9}{2}\frac{\xi\kappa\phi\dot{\phi}}{\kappa\rho_o}
\Bigg[H+\frac{\xi\kappa\phi\dot{\phi}\big[\mathcal{A}+\gamma^2(1+\xi\kappa\phi^2)^2\big]}{(1+\xi\kappa\phi^2)\big[1+\gamma^2(1+\xi\kappa\phi^2)^2\big]}\Bigg]\nn\\
&&-\frac{3\gamma^2(1+\xi\kappa\phi^2)^2(\mathcal{A} -\mathcal{A}^2)}{1+\gamma^2(1+\xi\kappa\phi^2)^2},\label{nmsB}\\
\beta\big(\phi,\dot{\phi}\big)&=&\Big(1+2\gamma^2(1+\xi\kappa\phi^2)^2\Big)\frac{\gamma^2(1+\xi\kappa\phi^2)^2+\mathcal{A}}{1+\gamma^2(1+\xi\kappa\phi^2)^2}-2\gamma^2(1+\xi\kappa\phi^2)^2.
\nn\\\label{nmsC}
\ea

In the low-energy limit with $\mathcal{A}\rightarrow1$, the above equations of motion reduce to the classical equations of motion. In the low-energy limit with $\mathcal{A}\rightarrow-1$, we obtain $H^2\rightarrow\frac{\kappa}{3}\frac{\gamma^2\rho_o}{(1+\gamma^2)^2}$, which corresponds to an expanding or a contracting de Sitter universe.

Since $\rho\!\geq0$,  from equation (\ref{nmsconstA}) we get $|\mathcal{A}|\leq1$ and
\ba
\rho\leq\frac{\rho_o}{4(1+\gamma^2)},\quad
(6\xi^2+\xi)\kappa\phi^2\dot{\phi}^2<\frac{\rho_o}{2(1+\gamma^2)},\quad\big(1+\xi\kappa\phi^2\big)\phi^4\leq\frac{\rho_o}{1+\gamma^2}. \nn\\
\label{inequala}
\ea
Substituting these inequalities into equations (\ref{nmsFra}), (\ref{nmsKGa}) and (\ref{nmsRea}), we obtain
\ba
&&|H|<\Bigg(\frac{\gamma}{\sqrt{3}}D^2(\gamma)+\frac{\sqrt{2}}{2}D(\gamma)C(\xi)\Bigg)\sqrt{\kappa\rho_o},\label{abHa}\\
&&\Big|\dot{H}\Big|<\Bigg[\frac{25+12\xi}{6}D^2(\gamma)
+\Bigg(6\sqrt{\frac{3}{2}}+\frac{(3\sqrt{3}+2)\gamma}{4\sqrt{6}}\Bigg)D^{\frac{3}{2}}(\gamma)C(\xi)\nn\\
&&\qquad\quad+\frac{5}{2}D^{2}(\gamma)C^2(\xi)\Bigg]\kappa\rho_o,\label{abdotHa}
\ea
where $C(\xi)\equiv\sqrt{\frac{\xi^2}{6\xi^2+\xi}}$ and $D(\gamma)\equiv\sqrt{\frac{1}{1+\gamma^2}}$. Equations (\ref{inequala}), (\ref{abHa}) and (\ref{abdotHa}) imply that the effective energy density, the Hubble parameter and the spacetime curvature remain finite in the effective theory of the modified STT-curvature quantisation.

In equation (\ref{nmsFra}), we define
\ba
R\big(\phi,\dot{\phi},\mathcal{A}\big)\equiv\frac{\rho}{(1+\xi\kappa\phi^2)^2}
+\frac{\gamma^2\rho_o}{2\big[1+\gamma^2(1+\xi\kappa\phi^2)^2\big]}
\Bigg(\frac{1-\mathcal{A}}{1+\gamma^2(1+\xi\kappa\phi^2)^2}-\frac{2\rho}{\rho_o}\Bigg),\nn\\\label{Rnms}
\ea
the Hubble parameter can be expressed by
\ba
H_{1,2}=\pm\sqrt{\frac{\kappa}{3}}\sqrt{R\big(\phi,\dot{\phi},\mathcal{A}\big)}\mathcal{A}
-\frac{\xi\kappa\phi\dot{\phi}}{1+\xi\kappa\phi^2}
\frac{\mathcal{A}+\gamma^2(1+\xi\kappa\phi^2)^2}{1+\gamma^2(1+\xi\kappa\phi^2)^2},\label{nmsaH1}
\ea
in which $H_1$ corresponds to the plus sign and $H_2$ corresponds to the minus sign. As mentioned in subsection \ref{subsec3b}, the branch  $H_{1}(t)$ describes a contracting quantum de Sitter universe in the limit $\mathcal{A}\rightarrow-1$ and an expanding classical universe in the limit $\mathcal{A}\rightarrow1$, whereas the  branch $H_{2}(t)$ describes a contracting classical  universe in the limit $\mathcal{A}\rightarrow1$ and an expanding quantum de Sitter universe in the limit $\mathcal{A}\rightarrow-1$.

\begin{figure}[!h]
\centering
\subfloat[\label{fig3a}
]{\includegraphics[height=4cm]{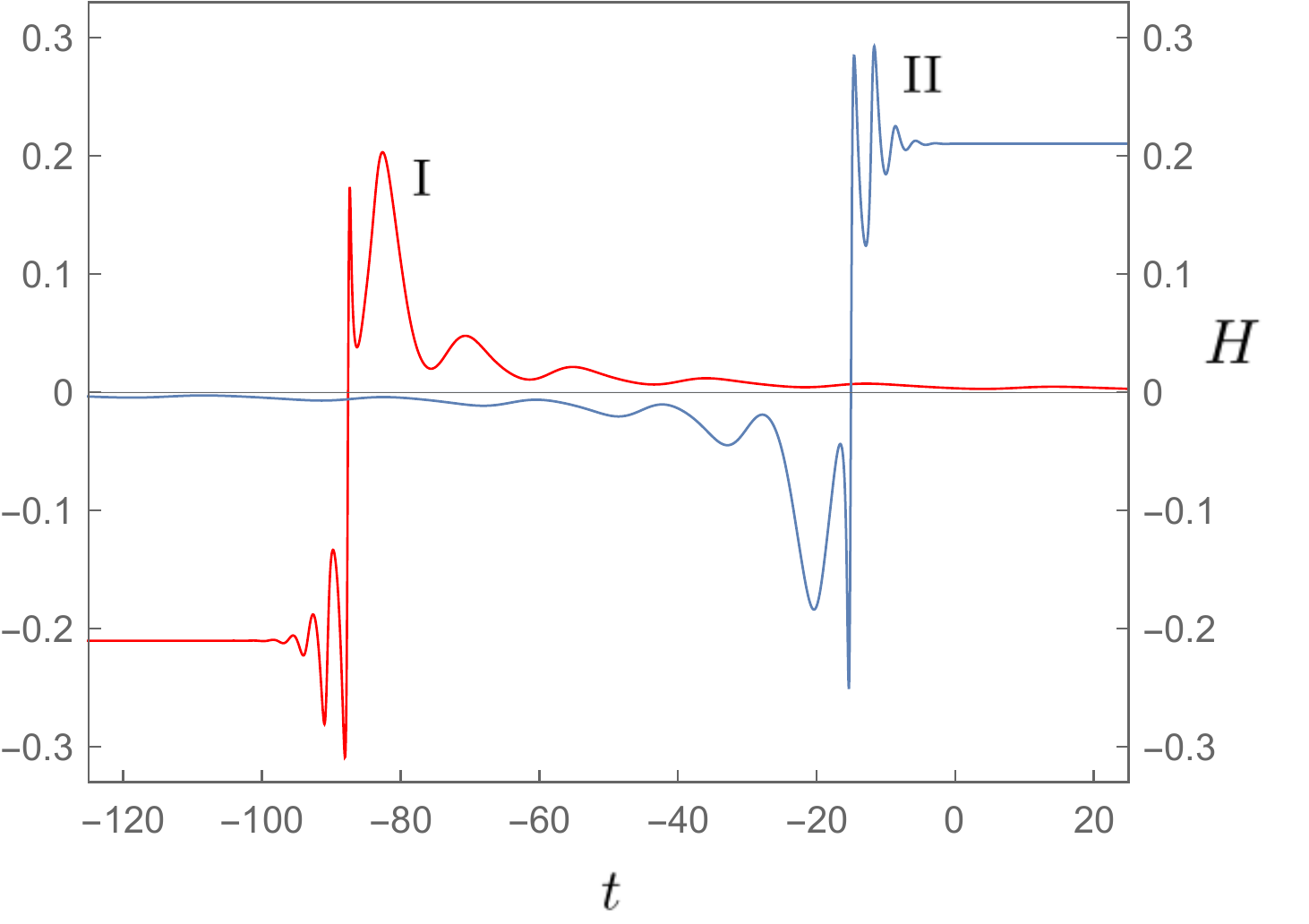}}
\hspace{1cm}
\subfloat[\label{fig3b}]{\includegraphics[height=4cm]{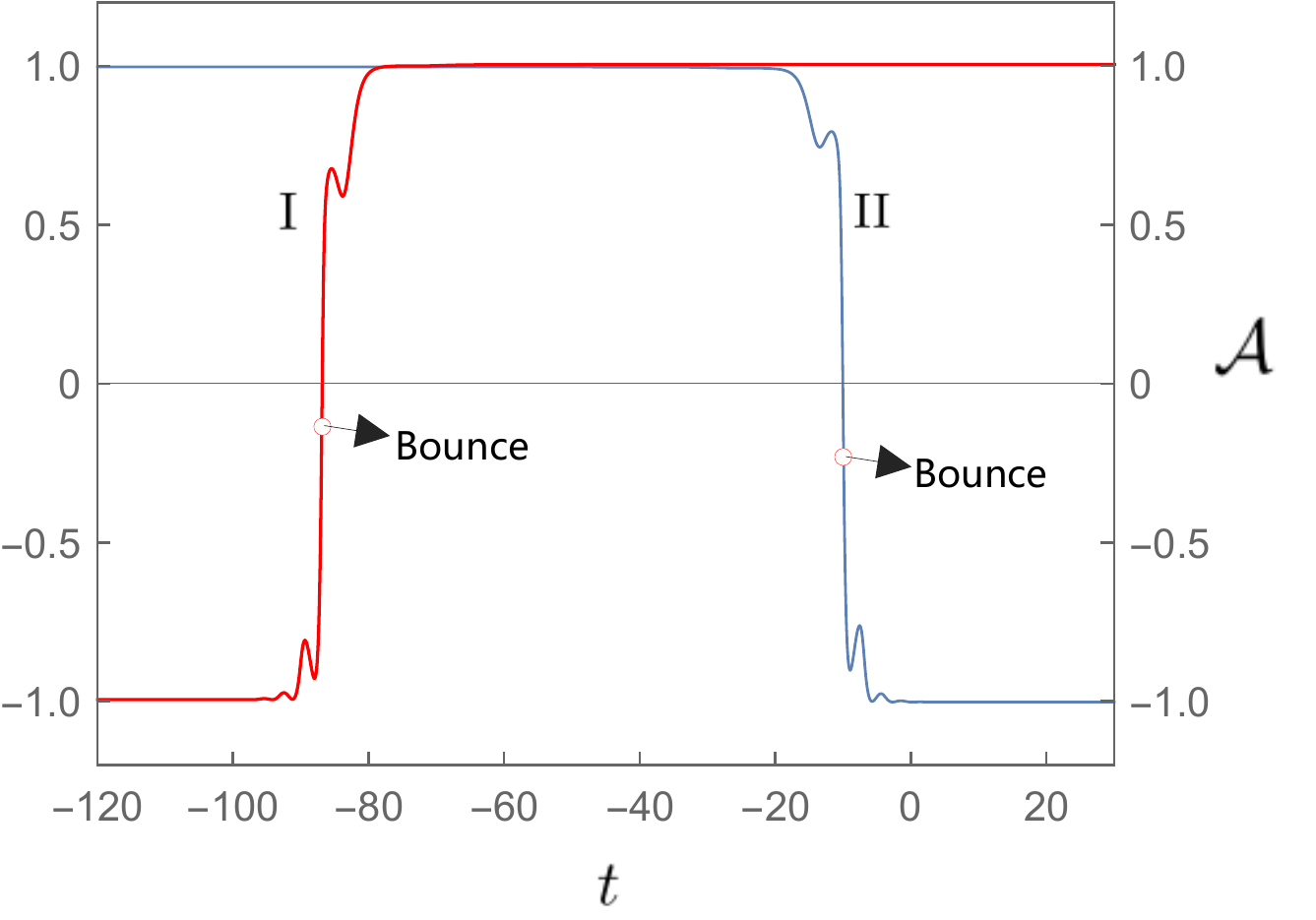}}
\captionsetup{justification=raggedright}
\caption{In Figure \ref{fig3a}, curve I shows the evolution of the Hubble parameter in the branch $H=H_1$ with the initial condition $\phi_i=0.05\kappa^{-\frac{1}{2}}$, $\dot{\phi}_i=-6.91\times10^{-6}\kappa$, $\mathcal{A}_i\rightarrow 1$, and curve II shows  the evolution of the Hubble parameter in the branch $H=H_2$ with the initial condition $\phi_i=0.05\kappa^{-\frac{1}{2}}$, $\dot{\phi}_i=-6.91\times10^{-6}\kappa$, $\mathcal{A}_i\rightarrow -1$.  For convenience of illustration, we set $\rho_o=\frac{3}{\Delta\kappa\gamma^2}\times10^{-3}$.  Figure \ref{fig3b} shows the corresponding evolution of $\mathcal{A}$ for curve I and II respectively, in which the hollow dot denotes the quantum bounce.}
\label{fig3}
\end{figure}

Numerical results show that under different initial conditions $\mathcal{A}$ always evolves from $-1$ to $1$ with respect to the proper time in the branch $H=H_{1}$ and evolves from $1$ to $-1$ with respect to the proper time in the branch $H=H_{2}$, which imply that in the branch $H=H_{1}$ an expanding classical universe is always born out of a contracting quantum de Sitter universe in the remote past and a  contracting classical universe will always undergo the bounce and evolve into an expanding quantum de Sitter universe in the asymptotic future (see for instance Figure \ref{fig3}).

\begin{center}
\begin{table}[h]
\begin{tabular}{|c|c|c|c|c|c|}\hline
  $\theta_i$&$t_{b}$&$\rho_{b}/\rho_o$& $\phi_{b}$ & $\dot{\phi}_{b}$&$\mathcal{P}(\phi_b)/\rho_{b}$  \\
  \hline
  $0$&$-45.1$&$8.76\times10^{-2}$ & $-1.73\times10^{-4}$ & $-0.981$&$1.29\times10^{-18}$ \\
  \hline
  $\pi/6$&$-5513.9$&$8.53\times10^{-2}$ & $-1.77\times10^{-4}$ & $-0.949$&$1.44\times10^{-18}$ \\
  \hline
  $\pi/3$&$-10682.8$&$7.81\times10^{-2}$ & $-1.88\times10^{-4}$ & $-0.854$ &$2.01\times10^{-18}$ \\
  \hline
  $\pi/2$&$-35838.9$&$6.68\times10^{-2}$ & $-2.09\times10^{-4}$ & $-0.711$&$3.61\times10^{-18}$ \\
  \hline
  $2\pi/3$&$-7242.1$&$2.33\times10^{-1}$ & $2.83\times10^{-3}$ & $-0.099$&$3.88\times10^{-14}$ \\
  \hline
  $5\pi/6$&$-3583.3$&$1.34\times10^{-1}$ & $-1.30\times10^{-2}$ & $0.016$&$9.86\times10^{-11}$ \\\hline
  \end{tabular}
\captionsetup{justification=raggedright}
\caption{ Relations between the initial conditions and values of parameters at the bounce in the effective theory of the modified STT-curvature quantisation.}\label{table2}
\end{table}
\end{center}

From Table \ref{table2}, we find that for almost all solutions with different initial conditions the energy densities at the bounce are extremely dominated by the kinetic energy density, which are similar as the cases in the effective theory of the STT-curvature quantisation. This fact can also be explained using arguments similar as in the last subsection. A slightly different thing from the case in the last subsection is that owing to the appearance of $\rho_o$ in the last term of equation (\ref{nmsKGa}) the maximum value that $\phi$ can reach during the evolution is seriously suppressed in the high energy density regime, which also contributes to the domination of the kinetic energy density at the bounce. Moreover, note that $\rho_b$ is much smaller than $\rho_o$  because $4(1+\gamma^2(1+\xi\kappa\phi_b^2)^2)\rho_b<\rho_o$.

\subsection{Effective dynamics from the STT-connection quantisation}
Using equations (\ref{qFr3})$\sim$(\ref{qRe3}),
we find that the effective Friedmann equation, Klein-Gordon equation of the STT-connection quantisation of the non-minimally coupled model are given by
\ba
&&\Bigg(H+\frac{\xi\kappa\phi\dot{\phi}}{1+\xi\kappa\phi^2}\cos b\Bigg)^2
=\frac{\kappa}{3}\frac{\rho}{(1+\xi\kappa\phi^2)^2}\Bigg(1-\frac{\rho}{\rho_o}\Bigg),\label{nmsFro}\\
&&\ddot{\phi}+3H\dot{\phi}+\frac{(6\xi^2+\xi)\kappa\phi\dot{\phi}^2+\frac{3}{2}(1-\cos b)\lambda\xi\kappa\phi^5
+\lambda\phi^3}{1+(6\xi^2+\xi)\kappa\phi^2}=0,\label{nmsKGo}
\ea
in which $\cos b$ is subject to the following constraint,
\ba
&&\cos^2b=1-\frac{\rho}{\rho_o}. \label{qconsto}
\ea
The evolution equation of $\cos b$ and the effective Raychadhuri equation read
\ba
\cos b\dot{\cos b}&=&\frac{3}{4\rho_o}
\Bigg(H\big(1+(6\xi^2+\xi)\kappa\phi^2\big)\dot{\phi}^2
-2\xi\lambda\kappa\phi^5\dot{\phi}\cos b\Bigg),\label{nmsoeomcos}\\
\dot{H}\cos b&=&
H\Bigg[\frac{3}{2}\Bigg(1-\frac{3\rho}{\rho_o}
+\frac{\lambda(1+\xi\kappa\phi^2)\phi^4}{4\rho_o}\Bigg)-\cos b\Bigg]\frac{2\xi\kappa\phi\dot{\phi}}{1+\xi\kappa\phi^2}\nn\\
&&-\frac{\cos b}{2(1+\xi\kappa\phi^2)}\Bigg[\kappa \dot{\phi}^2\Bigg(1-\frac{2\rho}{\rho_o}\Bigg)
+2\xi\kappa\big(\dot{\phi}^2+\phi\ddot{\phi}\big)\cos b\nn\\
&&\quad-\frac{3}{2}\frac{\lambda\xi^2\kappa^2\dot{\phi}^2\phi^6}{\rho_o}\Bigg].\label{nmsReo}
\ea
The above equations of motion reduce to the classical equations of motion in the low-energy limit with $\cos b\rightarrow1$, whereas in the low-energy limit with $\cos b\rightarrow-1$ the above effective equations describe a non-classical low energy universe. During the evolution, the Hubble parameter can vanish only when $\cos b=0$.

Since $\rho\leq\rho_o$,
it is not difficult to show that
\ba
&&|H|<\Bigg(\frac{1}{4\sqrt{3}}+\sqrt{2}C(\xi)\Bigg)\sqrt{\kappa\rho_o},\label{abHc}\\
&&\Big|\dot{H}\Big|<\Bigg[1+12\xi+\Bigg(8\sqrt{\frac{2}{3}}+\frac{1}{2}\sqrt{\frac{3}{2}}\Bigg)C(\xi)
+28C^2(\xi)\Bigg]\kappa\rho_o,\label{abdotHc}
\ea
where $C(\xi)\equiv\sqrt{\frac{\xi^2}{6\xi^2+\xi}}$. Hence, the spacetime curvature does not diverge for finite $\xi$ in the effective theory of the STT-connection quantisation.

From equation (\ref{nmsFro}), we find that the Hubble parameter can be expressed by
\ba
H_{1,2}=\Bigg(\pm\sqrt{\frac{\kappa}{3}}\frac{\sqrt{\rho}}{1+\xi\kappa\phi^2}
-\frac{\xi\kappa\phi\dot{\phi}}{1+\xi\kappa\phi^2}\Bigg)\cos b,
\ea
in which $H_1$ corresponds to the plus sign and $H_2$ corresponds to the minus sign. The branch $H=H_{1}$ can describe a contracting non-classical low energy universe in the limit $\cos b\rightarrow-1$ and an expanding classical universe in the limit $\cos b\rightarrow1$, and the branch $H=H_{2}$ can describe a contracting classical universe in the limit $\cos b\rightarrow1$ and an expanding non-classical low energy universe in the limit $\cos b\rightarrow-1$.

\begin{figure}[!h]
\centering
\subfloat[\label{fig4a}
]{\includegraphics[height=4cm]{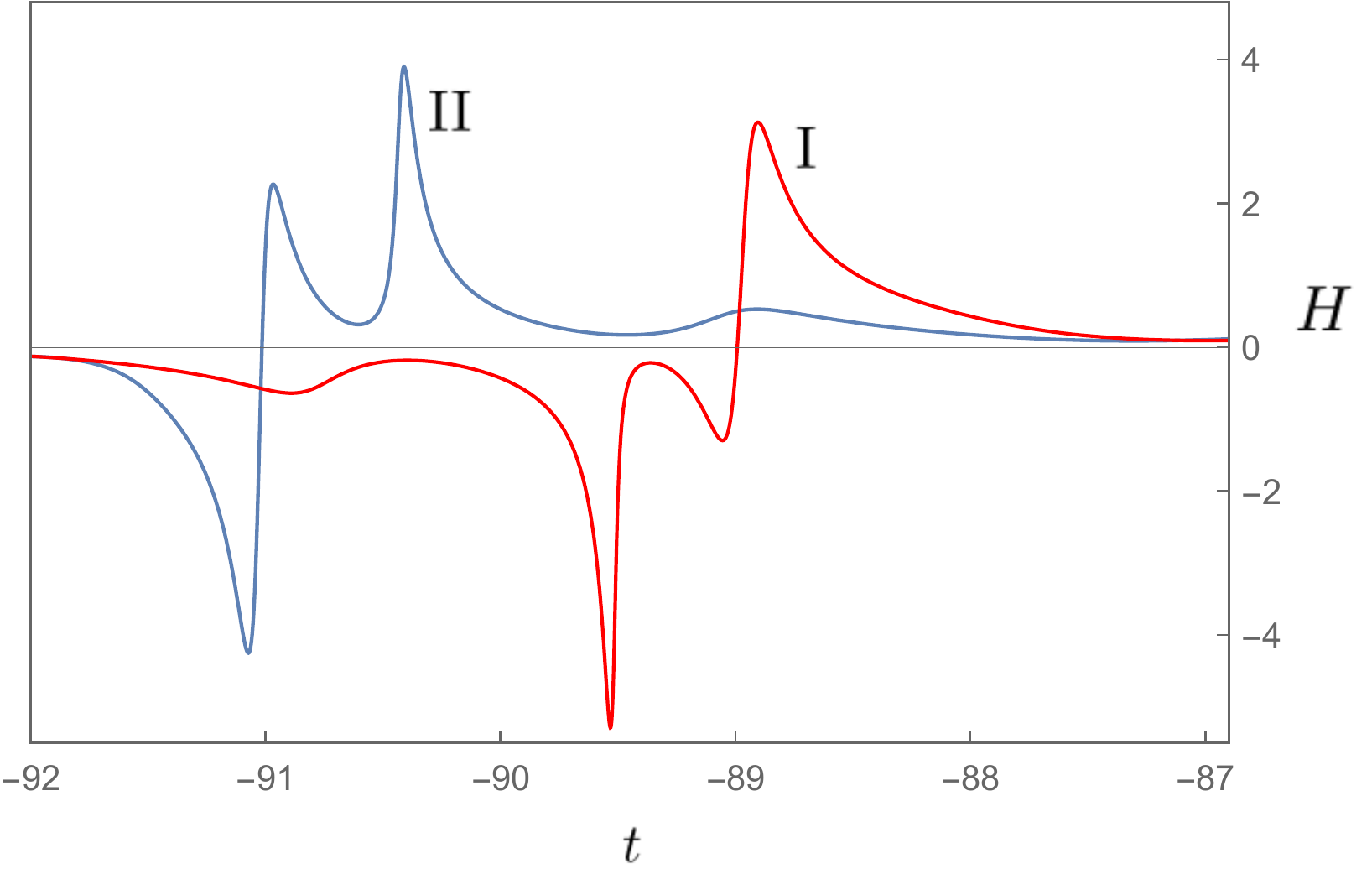}}
\hspace{1cm}
\subfloat[\label{fig4b}]{\includegraphics[height=4cm]{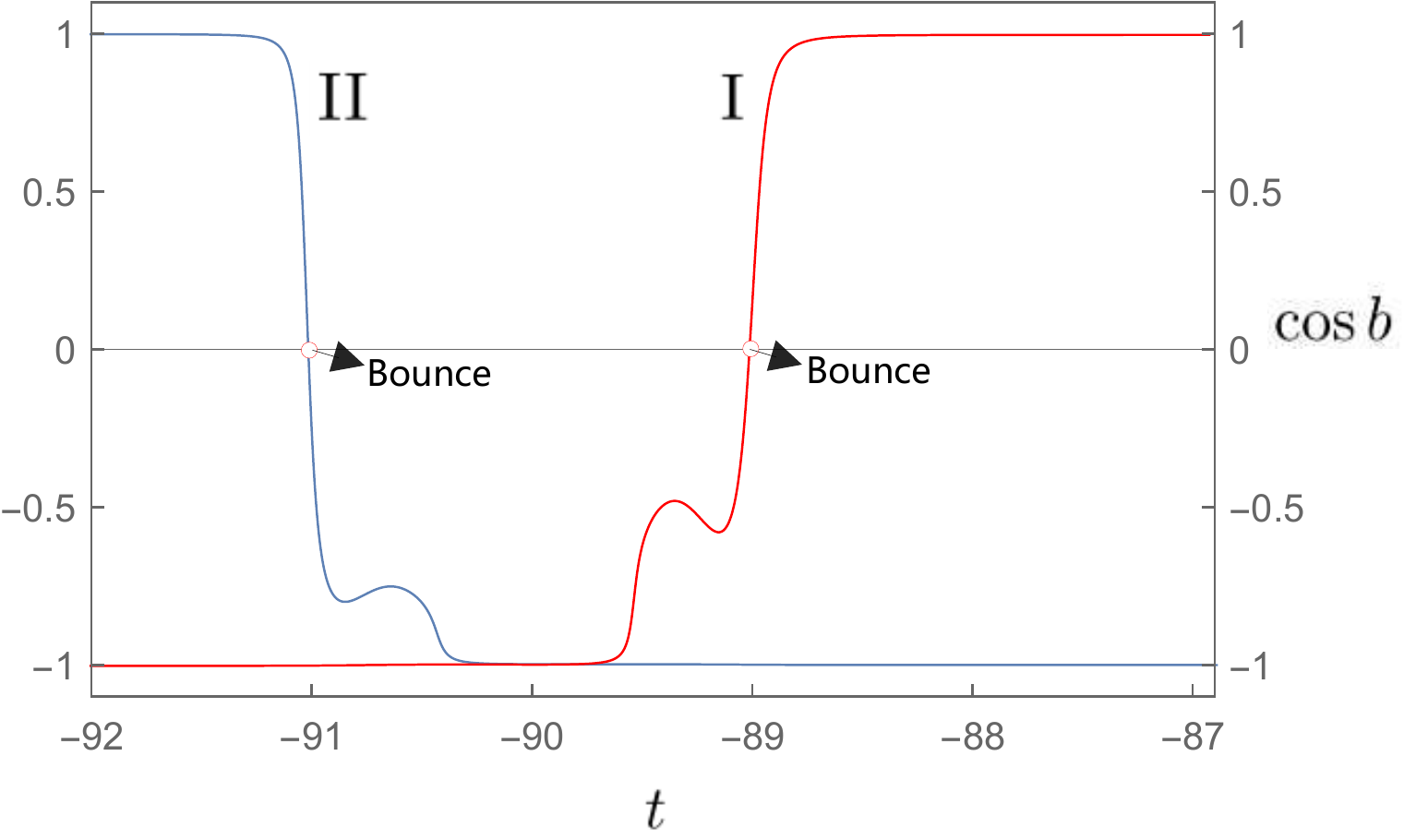}}
\captionsetup{justification=raggedright}
\caption{In Figure \ref{fig4a}, curve I shows  the evolution of the Hubble parameter in the branch $H=H_1$ with initial condition $\phi_i=0.065\kappa^{-\frac{1}{2}}$, $\dot{\phi}_i=-4.31\times10^{-6}\kappa$, $\cos b_i\rightarrow 1$ and  curve II shows  the evolution of the Hubble parameter in the branch $H=H_2$ with initial condition $\phi_i=0.065\kappa^{-\frac{1}{2}}$, $\dot{\phi}_i=-4.31\times10^{-6}\kappa$, $\cos b_i\rightarrow -1$.   Figure \ref{fig4b} shows the corresponding evolutions of $\cos b$ for curve I or II, in which the hollow dot denotes the quantum bounce.}
\label{fig4}
\end{figure}

 Numerical results show that  $\cos b$ always evolves from $-1$ to $1$ with respect to the proper time in the branch $H=H_{1}$ and from $1$ to $-1$ in the branch $H=H_{2}$,  which indicate that  in the effective theory an expanding classical universe is always born out of a non-classical low energy contracting universe in the remote past and a contracting classical universe will undergo the bounce and evolve into an expanding non-classical low energy universe in the asymptotic future (see for instance Figure \ref{fig4}). An interesting question is whether the current universe can be described by the expanding non-classical low energy universe. This possibility is discussed in detail in \cite{Han:2021}, where it is shown that there indeed  exist some models in which the slow-roll inflation predictions in a non-classical expanding universe with $\cos b\rightarrow-1$ can agree well with the current observations. However, using the arguments there, we find that the model we consider in this section cannot give spectral indices of slow-roll inflation compatible with the current observations on a non-classical low energy background.

\begin{center}
\begin{table}[h]
\begin{tabular}{|c|c|c|c|c|c|}\hline
  $\theta_i$&$t_{b}$&$\rho_{b}/\rho_o$& $\phi_{b}$ & $\dot{\phi}_{b}$&$\mathcal{P}(\phi_b)/\rho_{b}$  \\
  \hline
  $0$&$-43.2$&$1$ & $-2.93\times10^{-2}$ & $1.979\times10^{-2}$&$1.37\times10^{-9}$ \\
  \hline
  $\pi/6$&$-5511.8$&$1$ & $-2.97\times10^{-2}$ & $1.955\times10^{-2}$&$1.47\times10^{-9}$ \\
  \hline
  $\pi/3$&$-10681.1$&$1$ & $-3.08\times10^{-2}$ & $1.901\times10^{-2}$ &$1.84\times10^{-9}$ \\
  \hline
  $\pi/2$&$-36053.3$&$1$ & $-3.29\times10^{-2}$ & $1.761\times10^{-2}$&$2.72\times10^{-9}$ \\
  \hline
  $2\pi/3$&$-5720.4$&$1$ & $2.93\times10^{-1}$ & $-1.978\times10^{-3}$&$1.28\times10^{-3}$ \\
  \hline
  $5\pi/6$&$-2808.9$&$1$ & $2.36\times10^{-1}$ & $-2.462\times10^{-3}$&$3.44\times10^{-4}$ \\\hline
  \end{tabular}
\captionsetup{justification=raggedright}
\caption{Relations between the initial conditions and values of parameters at the bounce in the effective theory of the STT-connection quantisation.}\label{table3}
\end{table}
\end{center}

Comparing the data in Table \ref{table3} with those in Table \ref{table1}, we find that the numerical values of the parameters in the two tables are almost the same for each column, except for the ones in $\frac{\rho_b}{\rho_o}$.  This fact can be explained as follows. In the case $H=H_1$ and $\cos b>0$, the Klein-Gordon equation (\ref{nmsKGo}) can be written as
\ba
\dot{f}\big(\phi,\dot{\phi}\big)+\tilde{g}\big(\phi,\dot{\phi}\big)f\big(\phi,\dot{\phi}\big)
+\tilde{h}\big(\phi,\dot{\phi}\big)=0,\label{eomko}
\ea
where
\ba
f\big(\phi,\dot{\phi}\big)&\equiv&\sqrt{1+(6\xi^2+\xi)\kappa\phi^2}\dot{\phi},\nn\\
\tilde{g}\big(\phi,\dot{\phi}\big)&\equiv&\frac{\sqrt{3\kappa}}{1+\xi\kappa\phi^2}\Bigg[
\sqrt{\rho}-\sqrt{3\kappa}\xi\phi\dot{\phi}\Bigg]\sqrt{1-\frac{\rho}{\rho_o}},\nn\\
\tilde{h}\big(\phi,\dot{\phi}\big)&\equiv&\frac{\frac{3}{2}\Big(1-\sqrt{1-\frac{\rho}{\rho_o}}\Big)\lambda\xi\kappa\phi^5
+\lambda\phi^3}{\sqrt{1+(6\xi^2+\xi)\kappa\phi^2}}. \label{defnmsm}
\ea
Comparing equation (\ref{eomko}) with equation (\ref{eomk}), we find that
\ba
\tilde{g}-g=-\frac{3\kappa\xi\phi\dot{\phi}}{1+\xi\kappa\phi^2}
\Bigg[\sqrt{1-\frac{\rho}{\rho_o}}-\Bigg(1-\frac{\rho}{\rho_o}\Bigg)\Bigg].
\ea
 If the value of $\rho$ is not too large, we have $|\tilde{g}-g|\ll\!|g|$ for $\text{sgn}(\phi\dot{\phi})=-1$, in this case, the evolution of the scalar field in an expanding universe described by the branch $H=H_{1}$ in the effective theory of the STT-connection quantisation approach is similar as that in the effective theory of the STT-curvature quantisation.
\section{Conclusions and remarks}\label{sec5}
In this paper, we perform a preliminary study on the quantisation ambiguities and the related effective dynamics of STT in the Jordan frame in the $k=0$ FRW model.
 We quantise the Hamiltonian constraint of STT using three prescriptions, which we name as the STT-curvature quantisation, the modified STT-curvature quantisation, and the STT-connection quantisation, respectively. In the STT-curvature quantisation we quantise the Euclidean term and Lorentzian term in the same way while in the modified STT-curvature quantisation we quantise them differently, and the two approaches yield two different effective Hamiltonian constraints. In the STT-connection quantisation, whether the Euclidean term and the Lorentzian term are quantised in the same way or not only affect trivial details in the quantum operator but does not affect the effective Hamiltonian constraint. Hence, we obtain totally three different effective Hamiltonian constraints.

The effective equations of motion are derived from each Hamiltonian constraint and their implications are discussed respectively. Now, we briefly summarise the main features of the effective equations of motion of each quantisation approach.

In the effective theory followed from the STT-curvature quantisation, the evolution of the Hubble parameter is completely determined by the scalar field and its time derivative, in other words, given the initial values $(\phi_i,\dot{\phi}_i)$ in the contracting or expanding phase of the universe, the entire evolutions of the Hubble parameter and the scalar field can be fixed by the Klein-Gordon equation and the Raychadhuri equation, which are similar as the cases in the minimally coupled mainstream LQC. Nevertheless, the Hubble parameter does not vanish at a fixed energy density, which is unlike the case in the minimally coupled mainstream LQC.

In the effective theory from the modified STT-curvature quantisation, the effective equations are much more complicated than those from the STT-curvature quantisation approach. In particular,
the equations of motion involve an additional variable $\mathcal{A}$ which is related to the energy density of the scalar field through the constraint (\ref{qconst2}). It turns out that the sign of this variable is not a pure gauge choice but plays a crucial rule in the evolution of the universe. In the low-energy limit with $\mathcal{A}\rightarrow1$, the effective equations of motion reproduce the classical ones. In the low-energy limit with $\mathcal{A}\rightarrow-1$, the effective equations of motion describe an expanding or a contracting quantum de Sitter universe with an extremely large cosmological constant. In the effective dynamics, we have two physically different solutions of $H(t)$ for the same initial values of $\phi_i$, $\dot{\phi}_i$ and $\mathcal{A}_i$. The existence of two branches of solutions of $H(t)$ holds for both the non-minimally coupled case and the minimally coupled case. In the low-energy limit with $\mathcal{A}\rightarrow1$,  $H_{1}(t)$  describes an expanding classical universe while $H_{2}(t)$  describes an contracting classical universe, and in the low-energy limit with $\mathcal{A}\rightarrow-1$, $H_{1}(t)$ describes an expanding quantum de Sitter universe while $H_{2}(t)$ describes a contracting quantum de Sitter universe. In neither of the solutions can we obtain a classical universe in both the contracting phase and the expanding phase in the low-energy limit. Moreover, the Hubble parameter does not vanish at a fixed energy density either, which is like the case in the STT-curvature quantisation approach.

Like the case above, the equations of motion followed from the STT-connection quantisation also involve an additional variable $\cos b$ which is related to the scalar field through the constraint (\ref{qconst3}). In the low-energy limit with $\cos b\rightarrow1$, the effective equations of motion reproduce the classical ones while in the low-energy limit with $\cos b\rightarrow-1$ it is not the case. In the effective dynamics, we also obtain two branches of solutions of $H(t)$ for the same initial values of $\phi_i$, $\dot{\phi}_i$ and $\cos b_i$, and only one branch can describe the classical universe in the low-energy limit in the expanding phase. On the the hand, the Hubble parameter vanishes at a fixed energy density, which is unlike the cases in the STT-curvature and modified STT-curvature quantisation approaches. Moreover, the two branches of solutions of $H(t)$ converge in the minimally coupled case, which is unlike the case in the effective theory of the modified STT-curvature quantisation either.

To display the features of the effective dynamics of different quantisation approaches more explicitly,  we study a specific model with the non-minimal coupling function $F(\phi)=1+\xi\kappa\phi^2$ and self-interacting potential $V(\phi)=\frac{\lambda}{4}\phi^4$. It is shown that although  the big bang singularity is unavoidable in the classical case, in the effective theory of each quantisation approach the physical quantities always remain finite. Therefore, we can say that the absence of big bang singularity in the effective theory of LQC of STT in the Jordan frame is robust against quantisation ambiguities. Using numerical methods, we find some common features in the effective dynamics of each quantisation approach. For instance, the quantum bounce exists in the effective theory of each quantisation approach for various initial conditions; and the energy density at the bounce is extremely dominated by the kinetic energy density for almost all initial conditions, which has been explained semi-quantitatively.
Moreover, numerical results also show some features unique to the effective dynamics of each quantisation approach in this model.
In the effective dynamics from the STT-curvature quantisation, an expanding classical universe is always born out of a contracting classical universe via the quantum bounce, which is
similar as the cases in the minimally coupled mainstream LQC. In the effective dynamics from the modified STT-curvature quantisation, an expanding classical universe is always born out of a contracting quantum de Sitter universe in the remote past, which is described by the solution $H_1(t)$; and a contracting classical universe will always evolve  into an expanding quantum de Sitter universe in the asymptotic future, which is described by the solution $H_2(t)$; these cases resemble those in the minimally coupled modified LQC \cite{Li:2018b,Assanioussi:2019}. In the effective dynamics from the STT-connection quantisation approach, an expanding classical universe is also born out of a contracting non-classical low energy universe (but not the de Sitter one in the modified approach) in the remote past; and  a contracting classical universe will always evolve into an expanding non-classical low energy universe in the asymptotic future via the quantum bounce, which are quite different from the cases in the minimally coupled LQC. These numerical results concretely confirm our arguments in section \ref{sec3}.

Finally, let us make a few remarks at the end of this paper.

First, the quantisation prescriptions used for STT in this paper should be distinguished from the closed holonomy or open holonomy quantisation prescription for the minimally coupled LQC mentioned in section \ref{sec1}. In the minimally coupled LQC, the closed holonomy  quantisation or open holonomy quantisation basically refer to different prescriptions to promote the curvature to quantum operators by using closed holonomies or open ones. In the $k=0$ FRW model of the minimally coupled LQC, the two prescriptions are equivalent in the sense that they can yield the same quantum and effective Hamiltonian constraint. However, in the case of STT, besides the curvature term, a linear term of the connection variable is also involved in the Hamiltonian constraint, which is absent in the minimally coupled case. To quantise the curvature as well as the connection term, we use different prescriptions. In the STT-curvature quantisation, we first regularise the curvature using (\ref{Fabk}), then, we use the regularised curvature in (\ref{Fab}) and the identity in (\ref{Aai}) to regularise the connection variable and obtain (\ref{rAai}), hence, in this quantisation prescription the regularisation of the connection variable depends on the regularised curvature. In the STT-connection quantisation, we follow the opposite route by first regularising the connection using (\ref{opA}) and then use it to regularise the curvature, and it turns out that the effective Hamiltonian followed from the STT-connection quantisation are quite different from those followed from the STT-curvature quantisation.

Moreover, it should be mentioned that the identity in (\ref{Aai}) only holds for the $k=0$ FRW model, nevertheless, this identity can be generalised to some other spacetimes. For instance, in the $k=1$ FRW model, using the results in \cite{Ashtekar:2007}, it is not difficult to obtain
\ba
A_a^i
=-\frac{3}{4\kappa\gamma}\mathcal{V}^{\frac{1}{3}}_o\mathring{e}^b_j\epsilon^{ij}_{~~k}
\left\{p,F_{ab}^k\right\}+\frac{1}{r_o}\mathring{\omega}_a^i,\label{Aaiclosed}
\ea
in which $r_o$ denotes the radius (measured by the fiducial metric $\mathring{h}_{ab}$) of the fiducial 3-sphere. Obviously, the $k=0$ results can be recovered in the limit $r_o\rightarrow\infty$. By using (\ref{Aaiclosed}), we can extend the quantisation prescription in this paper to the $k=1$ model of STT, which will be left for future analysis.

The results in this paper have shown us that the effective dynamics of STT in the Jordan frame is more involved than the effective dynamics in the minimally coupled LQC. To distinguish which evolution is closer to the real evolution of our universe, more detailed studies on the background and perturbation dynamics in specific models of STT  are necessary in future works.

\ack
 The author thanks Dr. Long Chen for helpful discussions. This work is supported by NSFC (Grant No. 11905178) and Nanhu Scholars Program for Young Scholars of Xinyang Normal University.

\end{document}